\documentclass[11pt]{article}
 \usepackage{jheppub}
\pdfoutput=1
\usepackage{amsmath,amssymb}
\usepackage{epsfig}
\usepackage{hyperref}
\usepackage{color}
\usepackage{slashed}
\usepackage{placeins}

\usepackage{amsfonts}
\usepackage{graphicx, rotating}
\usepackage{epstopdf}
\usepackage{latexsym}
\usepackage{rotating}

\usepackage{subcaption}

\usepackage[normalem]{ulem}

\usepackage[font={small}]{caption}   % Writes captions of figures in small font

\definecolor{myred}{rgb}{0.7, 0, 0}
\definecolor{myblue}{rgb}{0, 0, 0.7}
\definecolor{mygreen}{rgb}{0.04, 0.7, 0.5}
\definecolor{mygray}{rgb}{0.1, 0.1, 0.1}

\hypersetup{colorlinks,citecolor=myred,linkcolor=myblue,urlcolor=myblue,linktocpage=true}

 \def\be   {\begin{equation}}   \def\ee   {\end{equation}}
 \def\ba   {\begin{array}}      \def\ea   {\end{array}}
 \def\bea  {\begin{eqnarray}}   \def\eea  {\end{eqnarray}}
 \def\bean {\begin{eqnarray*}}  \def\eean {\end{eqnarray*}}
 
 \def\bry{\begin{array}}
 \def\ery{\end{array}}

\numberwithin{equation}{section}

\begin{document}

\begin{flushright}
\footnotesize
DESY-23-103 \\
\end{flushright}
\color{black}

\title{
%\bf \LARGE
High-Temperature Electroweak Baryogenesis with Composite Higgs
}
\date{\today}

\author[a]{Benedict von Harling,}

\affiliation[a]{Institut de F\'isica d'Altes Energies (IFAE), The Barcelona Institute of Science and Technology, Campus UAB, 08193 Bellaterra (Barcelona), Spain}

\emailAdd{benedictvh@gmail.com}

\author[b]{Oleksii Matsedonskyi,}

\affiliation[b]{DAMTP, University of Cambridge, Wilberforce Road, Cambridge, CB3 0WA, United Kingdom}

\emailAdd{alexey.mtsd@gmail.com}

\author[c,d]{G\'eraldine Servant}

\affiliation[c]{Deutsches Elektronen-Synchrotron DESY, Notkestr. 85, 22607 Hamburg, Germany}
\affiliation[d]{II. Institute of Theoretical Physics, Universit\"{a}t Hamburg D-22761, Germany}

\emailAdd{geraldine.servant@desy.de}

\abstract{Electroweak Baryogenesis (EWBG) paired with the Composite Higgs (CH) scenario provides a well-motivated and testable framework for addressing the questions of the origin of the matter-antimatter asymmetry and the naturalness of the electroweak scale. The appeal of both concepts however experiences increasing pressure from the experimental side, as no conclusive signs of the corresponding new physics have been observed. In this note we present a modification of the minimal CH EWBG model, where electroweak symmetry breaking persists to temperatures far above the usually obtained upper bound of $\sim 100$~GeV. This allows for an increase of the mass of the main actor of EWBG in this scenario - the dilaton. Such a modification results in relaxing the tension with experimental data, generally modifying the phenomenology, and pointing at collider searches for the heavy dilaton as the main direction for its future tests.}

%
%\keywords{}
%

\maketitle

%%%%%%%%%%%%%%%%%%%%%%%%%%%%%%%%%%%%%%%%%%%%%%%%%%%%%%%%
%%%%%%%%%%    Main Text %%%%%%%%%%%%%%%%%%%%%%%%%%%%%%%%%%%%%%%%
%%%%%%%%%%%%%%%%%%%%%%%%%%%%%%%%%%%%%%%%%%%%%%%%%%%%%%%%

\section{Introduction}\label{sec:incon}

Electroweak Baryogenesis (EWBG)~\cite{Shaposhnikov:1987tw,Cohen:1990it} is one of the most popular mechanisms to explain the origin of the observed matter-antimatter asymmetry (see~\cite{Enomoto:2022rrl,Azatov:2022tii,Harigaya:2022ptp,Ellis:2022lft,Servant:2018xcs,vonHarling:2016vhf,Baldes:2016rqn,Cline:2021dkf,Postma:2022dbr,Cline:2021iff,Cline:2020jre,Kainulainen:2021oqs,Ellis:2019flb,Bruggisser:2017lhc} for recent studies and~\cite{Krauss:1999ng,Garcia-Bellido:1999xos,Konstandin:2011ds,Servant:2014bla,Hall:2019ank,Carena:2022qpf,Flores:2022oef} for alternative realizations). 
In this scenario, the baryon asymmetry is produced during a first-order electroweak phase transition (EWPT).
Such a first-order EWPT features expanding bubble walls which separate the phases of broken and unbroken EW symmetry. These walls are required to produce an asymmetry of left-handed particles and antiparticles on their opposite sides, due to CP-violating interactions.
The excess of left-handed states on one side of the walls is then precessed away by $(B+L)$-violating EW sphalerons, increasing the baryon number. A net baryon asymmetry, however, is only produced if on the other side of the walls the EW sphalerons are blocked. This  requires that the Higgs VEV $h$ is larger than the temperature $T$, $h/T \gtrsim 1$, in the EW-broken phase, meaning that the EWPT is strongly-first order. We refer the reader to Refs.~\cite{Cline:2006ts,Morrissey:2012db} for comprehensive  reviews of EWBG.

The two mentioned crucial ingredients of EWBG -- a strongly first-order EWPT, and sufficient CP violation -- are not provided by the Standard Model (SM), requiring new physics beyond it.  
Importantly, the requirement to have $h/T \gtrsim 1$ in the broken phase usually implies that the typical temperature of the phase transition does not exceed $100$~GeV. This is because thermal corrections typically decrease $h$ at large $T$ and therefore lead to small $h/T$. 
The upper bound on the temperature puts tight constraints on the new physics involved in EWBG. It has to be sufficiently light to be effective at $T\lesssim 100$~GeV, but also heavy enough to satisfy the experimental constraints on physics beyond the SM. 

These contradicting bounds can however be relaxed in scenarios of high-temperature EWBG, where the condition $h/T \gtrsim 1$ is ensured to hold at temperatures far above 100~GeV due to EW symmetry non-restoration (SNR)~\cite{Meade:2018saz,Baldes:2018nel,Glioti:2018roy,Matsedonskyi:2020mlz,Matsedonskyi:2020kuy,Matsedonskyi:2021hti,Matsedonskyi:2022btb,Biekotter:2022kgf,Chang:2022psj,Bai:2021hfb,Carena:2021onl}. In this paper we apply the EW SNR mechanism to EWBG in the composite Higgs framework based on the presence of extra singlet fermions.

We will consider the scenario where the EWPT is made first-order by linking it to the confinement phase transition of some new strong dynamics~\cite{Bruggisser:2018mus,Bruggisser:2018mrt,Bruggisser:2022rdm}. If the Higgs field is a composite state formed by this strong dynamics, it would be ``dissolved'' in the deconfined phase and only appear, with possibly a non-zero vacuum expectation value (VEV), after confinement takes place. Thus the EWPT can happen as a byproduct of confinement. Furthermore, if the confinement phase transition is first-order, the EWPT will have this property too, provided the temperature effects do not reduce the Higgs VEV to zero. Let us first analyse the temperature effects without SNR to justify the advantage of introducing the latter. 
In the presence of an approximate conformal symmetry in the strongly-interacting sector, which is the assumption we will make, the dynamics of the confinement phase transition is primarily controlled by the evolution of the dilaton field $\chi$, whose strength sets the scale of the condensate.
In this case, the reheat temperature after the confinement phase transition is related to the mass of the dilaton by
\be \label{eq:TRmchi}
T_r \propto \sqrt{m_\chi}\, ,
\ee
as will be explained in the following sections.
Requiring the reheat temperature $T_r$ to be less than $\sim 100$ GeV, to ensure $h/T>1$, one obtains the bound on the dilaton mass ${m_\chi \lesssim 500\;}$GeV~\cite{Bruggisser:2022rdm}, implying severe constraints from searches for new scalars at the LHC~\cite{Bruggisser:2022ofg}.

Furthermore, besides being responsible for a strong first-order phase transition, the dilaton can also source the required CP violation during the phase transition. For example, the phase of the top quark Yukawa coupling can vary across the bubble wall if it is linked to the value of the dilaton field,
\be\label{eq:mtsketch}
m_t \,\propto\, c_1+c_2 \, e^{i \phi} \chi ^\gamma \, .
\ee
This in turn induces CP-violating interactions between the top quarks and the bubble wall, which can result in a net baryon asymmetry~\cite{Bruggisser:2017lhc}. However, the dependence (\ref{eq:mtsketch}) also leads to CP-violating interactions between the top quark and the dilaton today, which contribute at two-loop order to the electron electric dipole moment (EDM) with a strength inversely proportional to the dilaton mass. The electron EDM is tightly constrained experimentally~\cite{ACME:2018yjb}, with estimates~\cite{Bruggisser:2022rdm} showing that it is only marginally consistent with the low dilaton mass region implied by Eq.~(\ref{eq:TRmchi}).

\begin{figure}
\begin{center}
\includegraphics[scale=0.5]{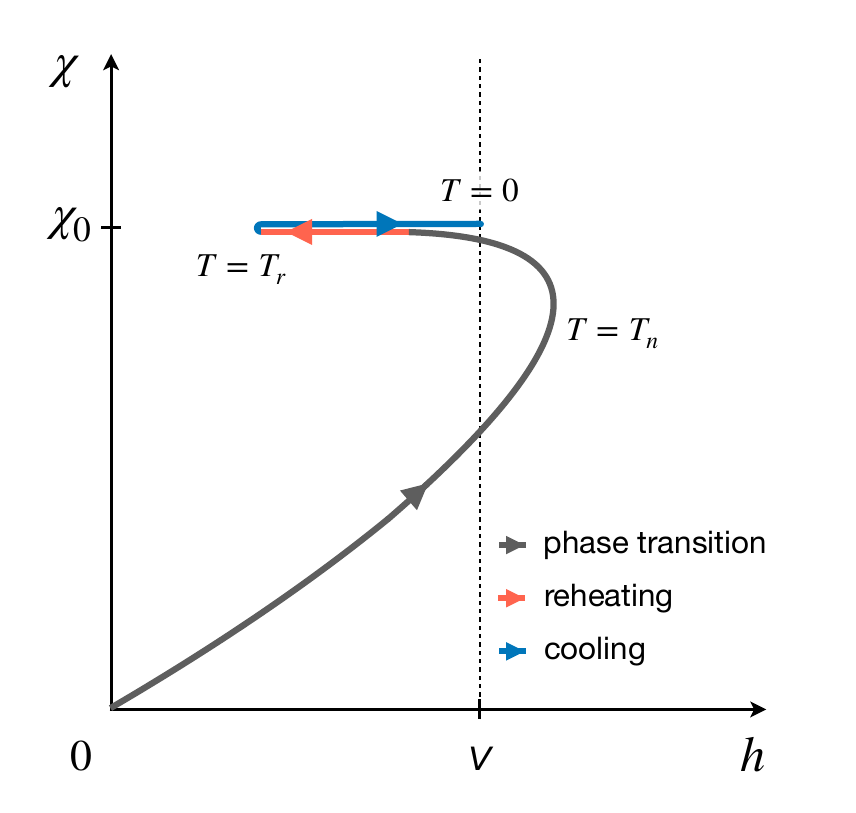}
\hspace{0.5cm}
\includegraphics[scale=0.5]{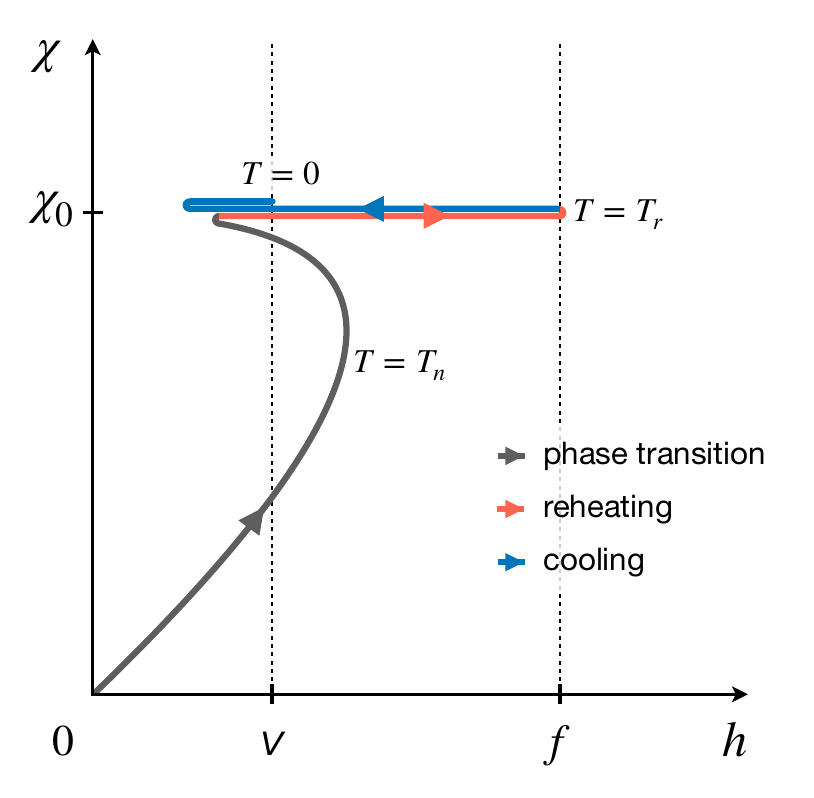}
\end{center}
\caption{{\it Schematic picture of the Higgs and dilaton evolution, for EWBG in the minimal composite Higgs model (left panel) and its 
high-temperature version introduced in this paper (right panel). At very high temperatures the system starts in the deconfined phase with $\chi=0$, $h=0$. At the nucleation temperature $T_n$ simultaneous confinement and EW phase transitions happen (grey lines), which are followed by reheating to the temperature $T_r$. 
In the minimal set-up the reheat temperature causes the Higgs VEV to decrease (red line, left panel), leading to a bound on the dilaton mass needed to maintain $h/T>1$.
The SNR mechanism instead ensures that $h$ grows after reheating (red line, right panel), allowing to maintain the condition
$h/T>1$ at any time after the phase transition. Note that here $h$ stands for $2m_W/g$ which differs from the composite Higgs parametrization used in the main text of the paper. }}
\label{fig:sketch}
\end{figure}

In this paper we will consider a modified scenario where we introduce new states which push the temperature of EW symmetry breaking above the typical reheat temperatures obtained for dilaton masses of order a few TeV. Hence even large dilaton masses would not cause the condition $h/T>1$ to be violated after reheating. The evolution of the Higgs and dilaton fields in such a scenario is shown in Fig.~\ref{fig:sketch}. High-temperature EWBG with a heavy dilaton implies a substantially different phenomenology, as we will discuss, and in particular relaxes the experimental constraints on the dilaton mass which we mentioned previously.

Our discussion focusses on minimal composite Higgs models with EWBG 
making use of
the dilaton field. Alternative possibilities were analysed e.g.~in Refs.~\cite{Grojean:2004xa,Bodeker:2004ws,Delaunay:2007wb,Grinstein:2008qi}  relying on new Higgs interactions, and in Refs.~\cite{Espinosa:2011eu,Chala:2016ykx,Xie:2020bkl,DeCurtis:2019rxl} using an additional composite Goldstone boson.
These alternative options can also be significantly affected by the presence of SNR, in particular potentially allowing for a larger mass of an additional Goldstone boson.

The paper is organized as follows. In Section~\ref{sec:Vh} we review how the EWPT can be linked to the confinement phase transition in the new strongly-interacting sector. In Section~\ref{sec:Vchi} we discuss the formalism to analyse confinement in the strong sector, and show that the properties of the latter (in particular, the dilaton mass) get severely constrained by the requirement of having broken EW symmetry after the transition. Sections~\ref{sec:Vh} and~\ref{sec:Vchi} are concise  summaries of Ref.~\cite{Bruggisser:2022rdm}. In Section~\ref{sec:snrferm} we introduce the source of high-temperature breaking of the EW symmetry, allowing to remove these constraints. The results of numerical scans of the parameter space are presented in Section~\ref{sec:num}. Sections~\ref{sec:cpv} and \ref{sec:grav} are respectively dedicated to the discussion of the electron EDM and gravitational waves in our model.

\section{Simultaneous Confinement and Electroweak Phase Transition}
\label{sec:Vh}

We would like to first review quantitatively how the EWPT can be linked to the confinement phase transition in the new strongly-interacting sector. We will assume the Higgs boson to be a composite state of new strong dynamics confining at the few-TeV scale. The relative lightness of the Higgs compared to other generic composite states is assumed to be the result of an approximate Goldstone-boson shift-symmetry. For concreteness, we consider the strong sector to possess a global $SO(5)$ symmetry which is spontaneously broken down to $SO(4)$ with the associated four Goldstone bosons forming the Higgs doublet~\cite{Agashe:2004rs}. The EW symmetry group of the SM is embedded into $SO(5)$ and can be either aligned with the unbroken $SO(4)$ or misaligned, in which case it is also broken.
The mass of the EW gauge bosons, which can be viewed as a proxy for the strength of  EW symmetry breaking, is given by
\be\label{eq:mw}
m_W^2 = \frac 1 4 g^2 f^2 \sin^2 \theta \, .
\ee
Here $f$ is the VEV of the $SO(5)\to SO(4)$ breaking condensate today, and $\theta$ measures how much the SM gauge group is misaligned with respect to the unbroken $SO(4)$. The canonically normalized Higgs field can then be chosen as $h=\theta f$. The overall strength of the symmetry breaking $f$ can be related to other mass scales of the composite sector, as we will discuss now. 

We will concentrate on the case where the composite sector is approximately conformally invariant in the UV. The conformal invariance is then spontaneously broken by the confinement and the resulting low-energy description of the model below the confinement scale should contain a relatively light state -- the dilaton $\chi$ -- the Goldstone boson of spontaneously-broken conformal invariance. 
Hence if the confinement phase transition happens at a temperature somewhat below the mass scale of other (generic) composite resonances, its dynamics should be well described by the dynamics of the dilaton transiting from $\chi=0$ to some non-zero value $\chi_0$ and thereby spontaneously breaking the conformal symmetry. The dilaton VEV today $\chi_0$ is crucial for both the properties of the phase transition and the phenomenology of the dilaton. It can be related to the scale $f$ as \cite{Bruggisser:2022ofg,Bruggisser:2022rdm}
\be\label{eq:chioff}
\chi_0= (g_*/g_\chi) f ,
\ee 
where $g_*$ is a typical coupling of meson-like states, and $g_\chi$ is the coupling of the dilaton. Assuming that the underlying strongly-interacting theory is an $SU(N)$ Yang-Mills, these couplings can be estimated as~\cite{Witten:1979kh}
\bea
g_* &=&c_k^{(h)} \frac {4\pi} {\sqrt N} \\ 
g_\chi &=& c_k^{(\chi)} \frac {4\pi} {N} \;  (\text{glueball}) \quad \text{or} \quad c_k^{(\chi)} \frac {4\pi} {\sqrt N} \;  (\text{meson})\, . \label{eq:gchi}
\eea
The two possible choices of $g_\chi$ correspond to the dilaton being either mostly a glueball-like or a meson-like state. The coefficients $c_k^{(h)}, c_k^{(\chi)}$ are some $N$-independent order-one numbers~\cite{Bruggisser:2022rdm,Baldes:2021aph}.
The lightest neutral state controlling the phase transition, the dilaton, is typically assumed to be glueball-like. This is in particular expected to be the case in strongly-coupled theories dual to 5D Randall-Sundrum models~\cite{Randall:1999ee} with Goldberger-Wise stabilisation~\cite{Goldberger:1999uk} (see also~\cite{Bellazzini:2013fga,Coradeschi:2013gda,Megias:2014iwa,Agashe:2019lhy,Pomarol:2019aae,CruzRojas:2023jhw} for 5D-based and~\cite{Chacko:2012sy,Aoki:2013qxa,LatKMI:2016xxi,Appelquist:2016viq,Appelquist:2022mjb} for 4D-based investigations of the emergence of a light dilaton).
We will also retain the option of a ``meson-like dilaton'' to keep our description more general, although we will show that this possibility is not phenomenologically viable in our set-up.

The relation~(\ref{eq:chioff}) links EW symmetry breaking to the dilaton VEV. Since the overall scale of the dilaton potential is much higher than that of the Higgs potential, the main properties of the phase transition are determined by the former. Hence one can speak of the EWPT being induced by the confinement phase transition, which we will look at more closely in the next section. 

Before moving to the discussion of the confinement phase transition, let us mention how the value of the misalignment angle $\theta$ in Eq.~(\ref{eq:mw}) is defined, which is also crucial for EW symmetry breaking. 
It is fixed by the Higgs potential which is in turn generated due to sources of explicit breaking of the $SO(5)$ symmetry~\cite{Panico:2015jxa} and approximately takes the form
\be\label{eq:vCHtree}
V_h = \alpha \sin^2 h/f + \beta \sin^4 h/f,
\ee
where $\alpha$ and $\beta$ are proportional to the $SO(5)$-breaking parameters. In the absence of breaking of conformal invariance, these coefficients have to scale with the dilaton VEV as 
\be
\alpha, \beta \, \propto \, (\chi/\chi_0)^4.
\ee
These two parameters have to be tuned~\cite{Matsedonskyi:2012ym,Redi:2012ha,Panico:2012uw} in order to reproduce the Higgs VEV $v\simeq 246\;$GeV in the minimum today while having a significantly larger $f\gtrsim 800\;$GeV, which is required by the experimental data~\cite{Grojean:2013qca}. We will assume $f =  800\;$GeV in the following. While in the previous study~\cite{Bruggisser:2022rdm} we have found $f$ to be bounded from above, $f \lesssim 1.1$~TeV, by a combination of the sphaleron washout and the collider bounds, this is not the case for the present scenario where the sphaleron washout bounds are removed, due to the SNR mechanism presented in Section~\ref{sec:snrferm}.
Furthermore, to account for the scaling of $f$ with $\chi$ in the arguments of the trigonometric functions we introduce the following kinetic term for the Higgs field
\be
{\cal L}_{\text{kin}} =  \frac 1 2 \frac{\chi^2}{\chi_0^2}(\partial_\mu h)^2.
\ee

\section{Properties of the Confinement Phase Transition}\label{sec:Vchi}

The properties of the confinement phase transition are determined  by the dilaton potential which we now discuss.
At zero temperature, the dilaton potential should contain a comformally-invariant quartic term and a source of explicit breaking of conformal invariance, which we denote as $\epsilon[\chi]$~\cite{cpr,Coradeschi:2013gda,Bellazzini:2013fga,Chacko:2012sy,Megias:2014iwa,Megias:2016jcw}:
\be\label{eq:vchi}
V_\chi = c_\chi g_\chi^2 \chi^4 - \epsilon[\chi] \chi^4\, .
\ee
The running of $\epsilon$ with the dilaton VEV is induced by the CFT dynamics and is determined by the RG equation
\be\label{eq:epsrunning}
\frac{\partial \epsilon}{\partial \log \mu} = \gamma_\epsilon \epsilon - c_\epsilon  \epsilon^2 /g_\chi^2\,,
\ee
{which we assume to be valid at least within the dilaton range $\chi \in (0,\chi_0)$ relevant for the phase transition.
In line with the assumption of approximate conformal invariance at high energies,} we take $\epsilon$ to be small and positive at large $\chi$, and $\gamma_\epsilon$ negative so that $\epsilon$ grows as $\chi$ decreases.
At $\chi=\chi_0$ the term $\propto \epsilon$ in Eq.~(\ref{eq:vchi}) equilibrates the scale-invariant quartic and thus produces a minimum in the dilaton potential. 

In the following, we will trade the parameter $\gamma_\epsilon$ for the dilaton mass $m_\chi$ as both are directly related. Indeed, neglecting possible Higgs-dilaton mixing for the moment, we have 
\be\label{eq:dilatonmass}
m_\chi^2 = -4 \gamma_\epsilon c_\chi  m_*^2 \, , 
\ee
where 
\be
m_* = g_\chi \chi_0
\ee
is the typical mass of the composite states.
We will  fix $c_\epsilon$ such as to minimize the value of $\epsilon$ at $\chi \ll \chi_0$, since the ratio $\epsilon/g_\chi^2$ controls the perturbative expansion of the dilaton potential and has to stay small.  
Finally, we will choose the remaining free parameter, $c_\chi$, somewhat less than one. Again this is to ensure that $\epsilon$ does not grow too much, since the corresponding term has to equilibrate the term $\propto c_\chi$ in the dilaton potential to form the minimum.

\begin{figure}
\begin{center}
\includegraphics[scale=0.6]{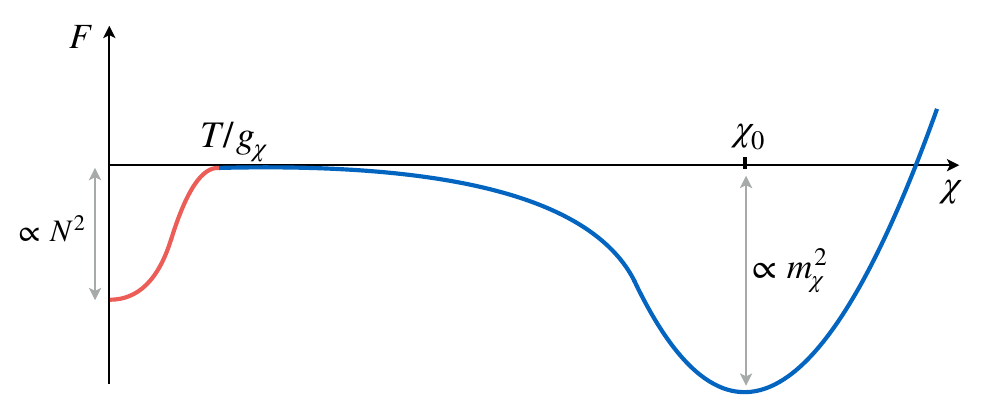}
\end{center}
\caption{{\it Sketch of the dilaton free energy at high temperatures. The red line shows the region which is dominated by the contribution of light CFT degrees of freedom. The blue line corresponds to the part which is dominated by the zero-temperature potential.}}
\label{fig:chipotential}
\end{figure}

We next discuss finite-temperature effects. For $\chi=0$, the theory is in the approximately  conformal, deconfined phase and the free energy is given by\footnote{For definiteness, we here use the result obtained for $\mathcal{N}=4$ $SU(N)$ super-Yang-Mills~\cite{Gubser:1996de}.}
\be\label{eq:fcft}
F_{\text{CFT}}[\chi=0]\, \simeq \, -\frac{\pi^2 N^2}{8} T^4\, .
\ee
As the dilaton VEV increases, the CFT eventually confines. For $\chi\gtrsim T/g_\chi$, the massive composite states decouple from the thermal bath and their contribution to the free energy vanishes. In this region, the free energy is well approximated by the zero-temperature potential~(\ref{eq:vchi}) discussed above. For $T/g_\chi \gtrsim \chi > 0$, on the other hand, the exact form of the free energy is not known since the theory is in the strongly-coupled regime. As discussed in more detail in Appendix~\ref{sec:temp}, we will model the free energy in this region by a smooth interpolation between the values $\chi=0$ and $\chi \sim T/g_\chi$.
A sketch of the dilaton free energy, including the zero-temperature part~(\ref{eq:vchi}) and the part in the conformal, deconfined phase~(\ref{eq:fcft}) is shown in Fig.~\ref{fig:chipotential}.

For the purpose of EWBG, it is crucial to ensure that the thermal effects do not drive the Higgs to small VEVs in the new phase, for the following reason. 
In EWBG, the baryon asymmetry is produced during the EWPT. CP-violating interactions between the bubble walls and the states in the surrounding plasma create CP asymmetries on both sides of the walls. The CP asymmetry in front of the bubble walls is subsequently turned into a baryon asymmetry due to the EW sphalerons which are active in that region since $h_{\text{SM}}\equiv(\chi/\chi_0)f \sin \theta < T$ (see Eqs.~(\ref{eq:mw}) and \eqref{eq:chioff} for the connection between the composite Higgs VEV and the SM Higgs VEV). Inside the bubbles, on the other hand, the EW sphalerons need to be inactive and thus $h_{\text{SM}}> T$, otherwise the overall asymmetry would be washed out. 
In the composite Higgs set-up, the 
criterion $h_{\rm SM}\gtrsim T$ is fulfilled for $T\lesssim 130\;$GeV, similarly to the SM \cite{Bruggisser:2022rdm}.
In particular, this bound needs to be satisfied by the nucleation temperature $T_n$, at which the phase transition takes place and the baryon asymmetry is generated. In addition, it also needs to be satisfied by the reheat temperature $T_r$ which is the temperature just after the phase transition has completed. The nucleation and reheat temperatures are generically different because of the latent heat which is released during a first-order phase transition and since the number of degrees of freedom in the deconfined and confined phase are very different. In our set-up, we always have $T_n <T_r$.
The reheat temperature can be derived from the conservation of the energy density before and after the phase transition,
\be\label{eq:econs}
\frac{\pi^2 g_{c}}{30} T_r^4 \, \simeq \, \Delta V  \, + \, \frac{3 \pi^2 N^2}{8} T_n^4  \, + \, \frac{\pi^2 g_d}{30} T_n^4 \,, 
\ee
where $g_{c}$ and $g_{d}$ are the numbers of SM (plus new fermions $S$, see next section) relativistic degrees of freedom in the confined and deconfined phase, respectively, and $\Delta V$ is the potential energy difference between the two minima. 
The terms proportional to the nucleation temperature $T_n$ are typically negligible compared to $\Delta V$. Using Eq.~(\ref{eq:vchi}) one finds that the latter is related to the dilaton mass via
\be\label{eq:dilatonmass1}
m_\chi^2  
= 16 \frac{g_\chi^2}{g_*^2} \frac{\Delta V}{f^2},
\ee
which allows to rewrite Eq.~(\ref{eq:econs}) as 
 \be
m_\chi^2 \simeq \frac{8\pi^2 g_c}{15}\frac{g_\chi^2}{g_*^2} \frac{T_r^4}{f^2}.
\ee
We see that the upper bound on the reheat temperature then implies an upper bound on the dilaton mass. Imposing the condition $T_r\lesssim 130$~GeV, we get
\be\label{eq:chimassbound}
m_\chi \lesssim \frac{g_\chi}{g_*} \, 500 \,\text{GeV},
\ee
which significantly squeezes the parameter space of the model. 
In particular, the bounds from collider searches for new spin-zero states (which can be interpreted as bounds on the dilaton~\cite{Bruggisser:2022ofg}) are very effective in this region and, together with other constraints, leave only a squeezed viable region~\cite{Bruggisser:2022rdm} on the border of sensitivity of the collider searches and the searches for the electron EDM. This can be seen in particular in the third plot of Fig.~\ref{fig:edm}.

In the next section we will introduce new fermions which allow to have $h_{\rm SM}/T>1$ at all reheat temperatures and thereby remove the upper bound on the dilaton mass~(\ref{eq:chimassbound}).

\begin{figure}
\begin{center}
\includegraphics[scale=0.9]{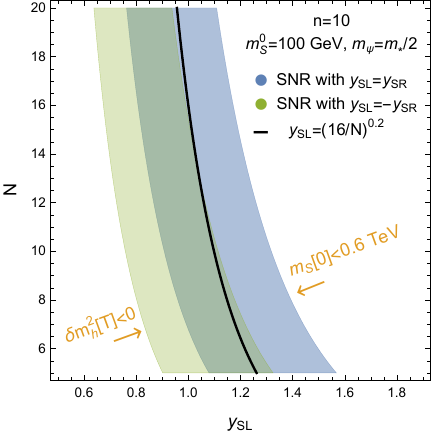}
\end{center}
\caption{{\it  Regions satisfying the SNR condition~(\ref{eq:snrcondition}) (left boundaries) and the bound on the fermion mass~(\ref{eq:snarmasscondition}) (right boundaries) for $y_{SL}=y_{SR}$ (blue region) and $y_{SL}=-y_{SR}$ (green region). The black line shows the dependence of $y_{SL}$ and $y_{SR}$ on $N$ which will be used in our numerical scans. The other parameters are set as specified in the plot.}}
\label{fig:snrprelim}
\end{figure}

\section{Introducing Electroweak Symmetry Non-Restoration}\label{sec:snrferm}

The main difference of this work with respect to the previous studies of EWBG in composite Higgs set-ups is the presence of new fermions, interacting with the Higgs boson in such a way that the EW symmetry remains broken even at high temperatures, thus removing the bound~(\ref{eq:chimassbound}). Concretely, we assume the presence of a new type of elementary SM-singlet Dirac fermions $S$ coupled to the composite sector through their composite partners $\psi$. 
Each of these come in $n$ copies. In principle, the presence of this type of fermions can be related to naturalness considerations for the Higgs mass~\cite{Matsedonskyi:2020kuy} or to dark matter~\cite{Matsedonskyi:2021hti}, but here we will not specify their origin and treat their couplings and masses as free parameters.
We assume the elementary fermions $S$ to couple to the strong sector as one component of the fundamental representation of $SO(5)$, neutral under the SM gauge group. In this way they break the $SO(5)$ explicitly and acquire non-derivative couplings to the Goldstone boson of the spontaneous $SO(5)\to SO(4)$ breaking -- the Higgs. 
The mass Lagrangian of the singlet fermions reads~\cite{Matsedonskyi:2020mlz}
\be
 {\cal L}_{\text{SNR}} \, = \, (g_\chi \chi/g_*) \, (y_{SL} \bar S_{L} \psi_R + y_{SR} \bar S_{R}  \psi_L \, + \, \text{h.c.}) \cos h/f \, - \, m_{\psi}^0 \bar \psi \psi - m_S^0 \bar S S\,,
\ee
where $y_{S L},y_{S R}$ are dimensionless mixing parameters. The fact that the interactions between the elementary and the composite states are now proportional to $\cos h/f$ follows from both types of fermions being SM singlets, which can have a mass mixing without breaking EW symmetry, i.e.~for $h=0$.
In the following, we will set the mass of the composite singlets to be $m_\psi^0 = c_\psi  g_\chi \chi$, and treat $c_\psi$ as a free parameter, with a value of order 1. The mass $m_S^0$ of the elementary singlets is instead independent of $\chi$.

\begin{figure}
\center
\includegraphics[scale=0.8]{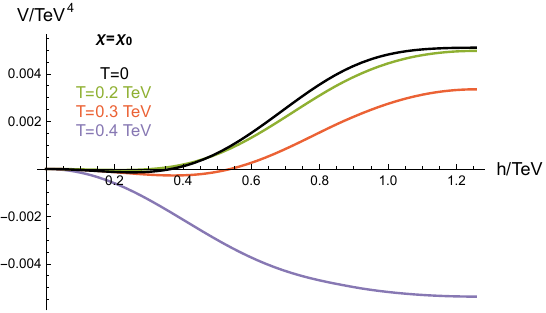}
\caption{{\it Higgs potential at various temperatures, for the glueball dilaton and $n=10$, $y_{SL}=1$, $m_{\chi}=1.5$~TeV, $N=15$, $c^{(\chi)}_k=1$, and the other parameters set as in Table~\ref{tab:bench}.}}
\label{fig:VhofT}
\end{figure}

The mass eigenvalues are approximately given by (assuming $y_{SL} f,y_{SR} f,m_S^0 \ll m_\psi^0$) 
\be\label{eq:chmasses}
m_S[h] \, \simeq \,  m_S^0 - \frac {y_{SL} y_{SR} f^2} {m_\psi^0} \cos[h/f]^2\,,\qquad
m_{\psi}[h] \, \simeq \, m_{\psi}^0 + \frac{(y_{SL}^2+y_{SR}^2)f^2}{2m_\psi^0} \cos[h/f]^2,
\ee
where we have set $(g_\chi \chi/g_*)=f$ for brevity.
At this point one can see how the lighter set of fermionic mass eigenstates can contribute to EW symmetry breaking at high temperature. Expanding for large temperatures, the thermal contribution of the new fermions to the Higgs mass is given by
\be
\delta m_h^2\, \simeq \, T^2 \frac{n}{12} (m_S^2[h])''_h  \,\simeq\, T^2 \frac{n}{3} \frac{y_{SL} \, y_{SR} \, m_S[0]}{m_{\psi}^0}
\ee
which can become negative for $y_{SL} \, y_{SR} \, m_S[0]/m_{\psi}^0 <0$. 
This negative contribution can overcome the positive thermal correction induced by the SM states
\be
\delta m_h^2[T]_{\text{SM}} \, \simeq \, 0.4 \,T^2
\ee
if
\be\label{eq:snrcondition}
-n y_{SL} \, y_{SR} \frac{m_S[0]}{m_\psi^{(0)}} \, \gtrsim \, 1.2.
\ee
As a result of having a negative mass, the Higgs VEV will be pushed to non-zero values even at high temperatures. The broken EW symmetry will ensure that sphalerons are inactive and the baryon asymmetry remains unaffected by them even when the reheat temperature is much larger than the EW scale. This does not mean that the asymmetry stays completely intact. Reheating to the temperature $T_r$ after the phase transition is generically expected to produce matter and antimatter in equal amounts. Hence the asymmetry generated in the plasma with temperature $T_n$ will be diluted by a factor of $(T_n/T_r)^3$. This effect has to be considered when computing the baryon asymmetry, however it is not as problematic as the exponential dilution of the asymmetry by the sphalerons which we got rid of.

\begin{table}[t]
\centering
\begin{tabular}{|c|c|c|c|c|c| c| c| c| c| c| c| c| c| c| c|}
\hline
$f$ & $c^{(h)}_k$ & $c_\epsilon$ & $c_\chi$   & $c_\alpha$ & $c_\beta$  & $\gamma_y$ & $c_y$ & $y_t[\chi_0]$  & $ c_\psi$ &  $m_S^0$\\
\hline
800 \text{GeV} & 1&  0.5 & 0.2 & -0.3 & 0.3  & -0.3 & 1.875 & $0.6 \sqrt{\lambda_t g_\star}$ & 0.5  & 100\;\text{GeV}\\
\hline 
\end{tabular}
\caption{\it \small Choice of parameters for our numerical studies. Here $c_\psi$ determines the mass of the composite new fermions via $m_\psi^0 = c_\psi  g_\chi \chi$, $\lambda_t$ is the top Yukawa and $y_t$ is one of two contributions to the right-handed top mixing. Furthermore, we set $y_{S L}=y_{S R}$ and choose $y_{S L}$ close to the parametrization given in Eq.~\eqref{eq:yofN}. We choose different values for the remaining parameters $n$, $m_\chi$, $N$, $c^{(\chi)}_k$ which we specify when we discuss the corresponding results.}
\label{tab:bench}
\end{table}

An extra condition for the new fermions to be effective at temperatures around $100$~GeV is that their mass is sufficiently small, otherwise their thermal population would be significantly suppressed. We have found numerically that for $n= {\cal O}(10)$ this approximately requires
\be\label{eq:snarmasscondition}
m_S[0]\lesssim 0.6~\text{TeV}.
\ee

In Fig.~\ref{fig:snrprelim} we show the regions satisfying the conditions~(\ref{eq:snrcondition}) and (\ref{eq:snarmasscondition}) in the $N-y_{S L}$ plane, fixing the other mixing of the new fermions as either $y_{S R}=y_{S L}$ or $y_{S R}=-y_{S L}$.
In our numerical scans we focus on the case $y_{SL}=y_{SR}$, and parametrize these mixings as a function of $N$ and $n$ with
\be\label{eq:yofN}
y_{S L} \, = \, y_{S R} \, \simeq \, (10/n)^{1/2} (16/N)^{1/5},
\ee
which corresponds to the black line in the center of the corresponding SNR region in Fig.~\ref{fig:snrprelim} (the exact value of the mixing in the scan will be adjusted to achieve SNR in each parameter space point).

In Fig.~\ref{fig:VhofT} we plot the thermal potential along the Higgs direction for $\chi=\chi_0$ and $\chi=\chi_0/2$ for different temperatures. As one can see, EW symmetry is broken even at high temperatures due to the presence of the SNR fermions. 

We should also mention that non-renormalizable Higgs interactions controlled by the scale $f$ produce higher-order thermal corrections growing as $\propto T/f$~\cite{Ahriche:2010kh}, and therefore our description is only valid at temperatures
\be
T \ll f.
\ee
In practice, the reheat temperatures in the regions of parameter space relevant for us never get close to this bound.

\section{Numerical Results for the Phase Transition} \label{sec:num}

\begin{figure}
\center
\includegraphics[scale=0.55]{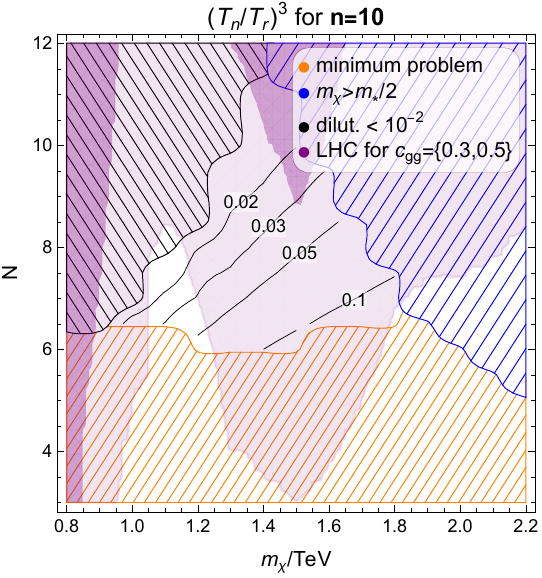}
\hspace{-0.1cm}
\includegraphics[scale=0.55]{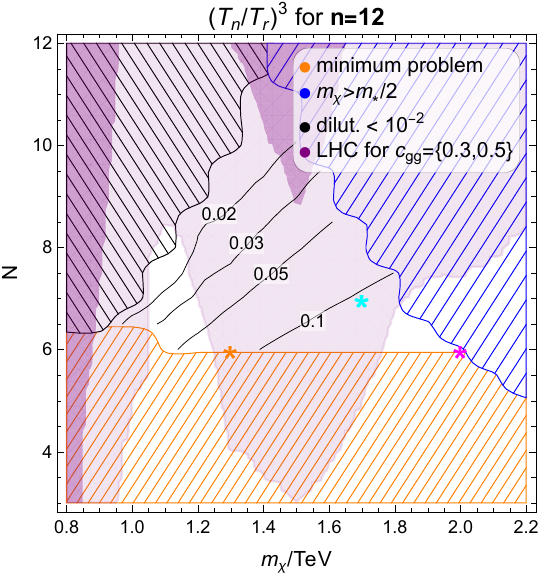}
\hspace{-0.1cm}
\includegraphics[scale=0.55]{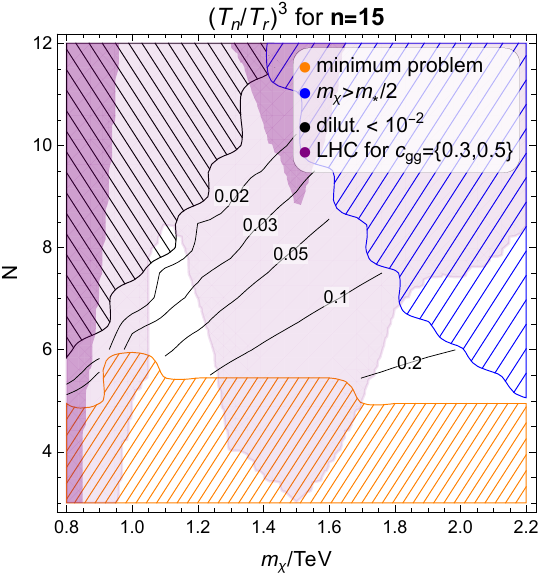} 
\caption{{\it Contours (in black) of the dilution factor of the baryon asymmetry and various constraints for the glueball-like dilaton and varying $n$. We set $c_k^{(\chi)}=1$ and scan over $y_{S L}$ as discussed in the text. In the orange-hashed region the Higgs potential develops a global minimum at a wrong Higgs VEV. In the grey-hashed region the entropy injection during reheating leads to the dilution factor of the baryon asymmetry being smaller than $10^{-2}$. In the blue-hashed region the validity of our effective low-energy description degrades due to the closeness of the cutoff physics. Two shades of purple show regions excluded by the LHC for $c_{gg}=0.3$ (darker shade) and $c_{gg}=0.5$ (lighter shade). The colored stars in the middle plot correspond to the benchmark points in Table~\ref{tab:bmp} used for the gravitational-wave spectra in Fig.~\ref{fig:gwspectra}.
}} 
\label{fig:scans_1}
\end{figure}

In this section we present results of numerical scans of our model parameter space. The scans are performed with parameters set as in Table~\ref{tab:bench}. The mixings of the new fermions are chosen close the parametrization in Eq.~(\ref{eq:yofN}) as discussed in more detail below. The resulting regions in the $N-m_\chi$ plane with potentially viable EWBG are shown as white regions in Fig.~\ref{fig:scans_1}, for three different choices of the number $n$ of new fermions $S$. We only analyse those properties relevant for EWBG which are directly related to the phase transition, leaving the CP violation and the baryon asymmetry aside for the moment.\footnote{We compute the phase transition properties using our custom code and the FindBounce package~\cite{Guada:2020xnz}.}

\begin{figure}
\center
\includegraphics[scale=0.55]{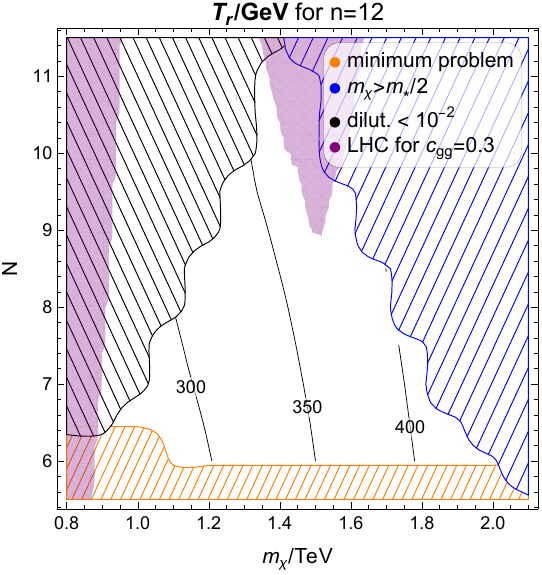}
\hspace{-0.1cm}
\includegraphics[scale=0.55]{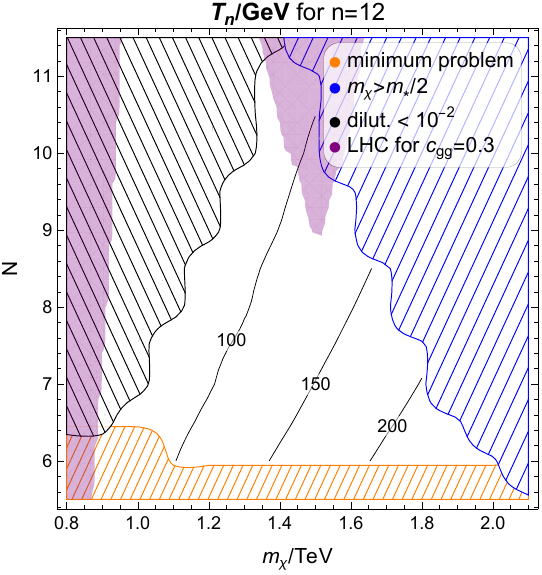}
\hspace{-0.1cm}
\includegraphics[scale=0.55]{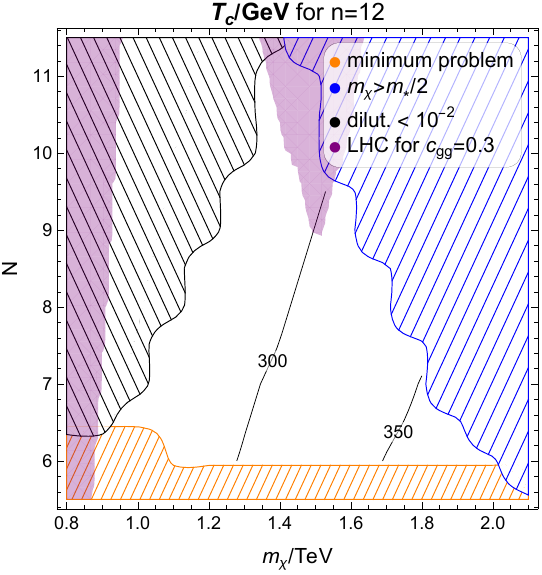} 
\caption{{\it Contours (in black) of the reheating, nucleation and critical temperatures (in GeV) for the glueball dilaton with parameters as chosen for Fig.~\ref{fig:scans_1} and $n=12$. The color code is the same as in Fig.~\ref{fig:scans_1}.
}}
\label{fig:scans_2}
\end{figure}

Let us discuss the results presented in Fig.~\ref{fig:scans_1} more closely. First of all, the overall dilaton mass scale is allowed to reach 2 TeV, far beyond the bound~(\ref{eq:chimassbound}). This is the result of the presence of the new fermions which make the EW sphalerons inactive even at reheat temperatures greater than $\sim 130$ GeV. A sufficiently large negative thermal correction to the Higgs mass generated by the new fermions is ensured by choosing correspondingly large values of $y_{SL}=y_{SR}$. During our scan we first check whether the benchmark choice of Eq.~(\ref{eq:yofN}) provides $h_{\text{SM}}/T > 1$ for all the relevant temperatures after the phase transition. If this is not fulfilled, we increase the values of  $y_{SL},y_{SR}$ in steps of $5\%$ until the condition $h_{\text{SM}}/T > 1$ is met, or the point becomes excluded because of the appearence of a wrong global minimum in the Higgs potential, which we discuss below.
We present in Fig.~\ref{fig:scans_2} contours of the critical temperature $T_c$ at which tunneling becomes energetically possible, the nucleation temperature $T_n$ at which the phase transition takes place, and the reheat temperature $T_r$ just after the phase transition.
Both $T_n$ and $T_r$ typically exceed $100$~GeV, without triggering sphaleron washout due to SNR.  In comparison, the viable baryogenesis region in minimal composite Higgs without extra singlets is associated with a nucleation temperature in the $5-80$ GeV range and a critical temperature of $100-160$ GeV, while the dilaton mass is in the $250-500$ GeV range  \cite{Bruggisser:2022rdm}.

The main constraint which does not allow even larger dilaton masses is the validity of our effective field theory description of the phase transition. When the dilaton mass $m_\chi$ becomes comparable to the typical mass of generic composite states $m_*$, the latter can not be integrated out in the way we did to make our computations tractable. We impose the bound $m_\chi < m_*/2$ (with a degree of arbitrariness) to ensure the validity of the effective field theory. The regions in the plots where this bound is not fulfilled are hashed in blue.

As we have mentioned previously, reheating after the phase transition reduces the baryon asymmetry by a factor $\sim (T_n/T_r)^3$. Contour lines of this dilution factor are shown in Fig.~\ref{fig:scans_1}. If the dilution factor is below $10^{-2}$, it is expected to be difficult to generate the needed amount of baryon asymmetry within the standard EWBG scenario.\footnote{We obtain the estimate $10^{-2}$ for the bound on the dilution factor based on~\cite{Bruggisser:2017lhc} and~\cite{Bruggisser:2018mrt}.} The regions in the plots where this bound is not fulfilled are hashed in grey.
Note that the gray area can be shifted towards larger $N$ if the parameter $c_k^{(\chi)}$ (defined in Eq.~(\ref{eq:gchi})) is increased. However, the same change in $c_k^{(\chi)}$ strengthens the collider constraints, which we discuss next, hence eventually there is no benefit in terms of increasing the viable parameter space.

\begin{figure}
\center
\includegraphics[scale=0.83]{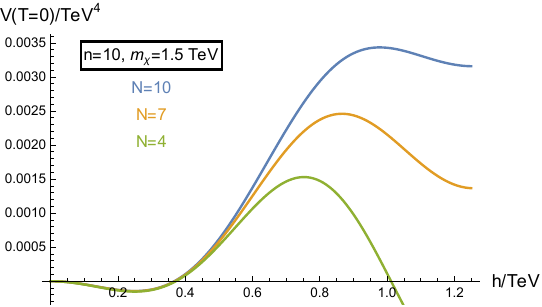}
\includegraphics[scale=0.83]{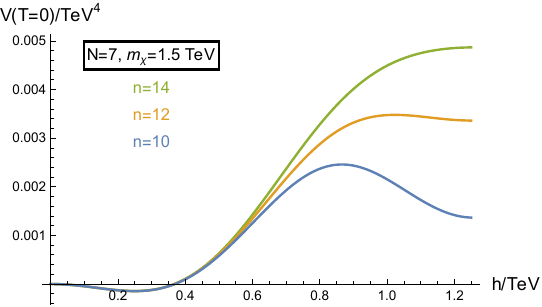}
\caption{{\it Zero-temperature Higgs potential for a glueball dilaton with $m_\chi = 1.5$~TeV and different choices of $N$ and $n$. 
We set $c_k^{(\chi)}=1$ and choose $y_{SL}$ according to Eq.~(\ref{eq:yofN}). The presence of a global minimum at large Higgs VEVs is responsible for the excluded hashed orange region in Fig.~\ref{fig:scans_1}.}}
\label{fig:hinstab}
\end{figure}

A sizeable part of the model parameter space 
is excluded due to the presence of a wrong deeper minimum in the Higgs potential around $h=\pi f/2$. The corresponding regions in the plots are hashed in orange.
This minimum is generated by the one-loop zero-temperature corrections induced by the new fermions. Since the one-loop correction has contributions proportional to $n \, m_{\psi,S}^4 \log m_{\psi,S}^2$, it is easy to show, using the expressions for the fermion masses~(\ref{eq:chmasses}) and the scaling~(\ref{eq:yofN}), that it decreases with growing $N$ and $n$. 
The dependence of the depth of the additional minimum on $N$ and $n$ is demonstrated in Fig.~\ref{fig:hinstab}. The constraint coming from the presence of the new global minimum is the only bound which is substantially sensitive to the number $n$ of new fermions as long as the SNR condition is fulfilled. Varying various parameters controlling the other bounds, we were not able to find viable parameter space for $n\lesssim 10$.

\begin{figure}
\center
\rotatebox{90}{\parbox[c]{3.5cm}{\centering \small glueball}}
\hspace{-0.25cm}
\includegraphics[width=0.33\textwidth]{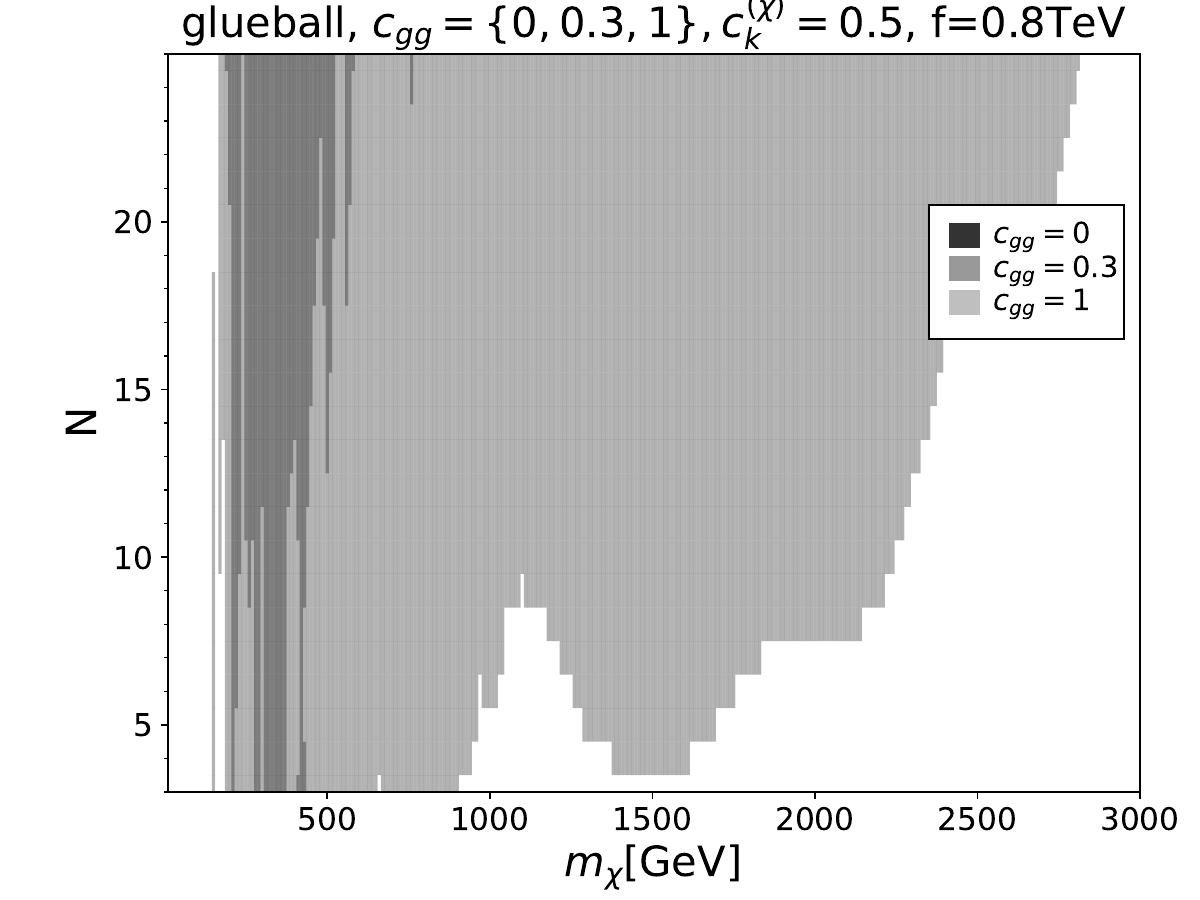}
\hspace{-0.5cm}
\includegraphics[width=0.33\textwidth]{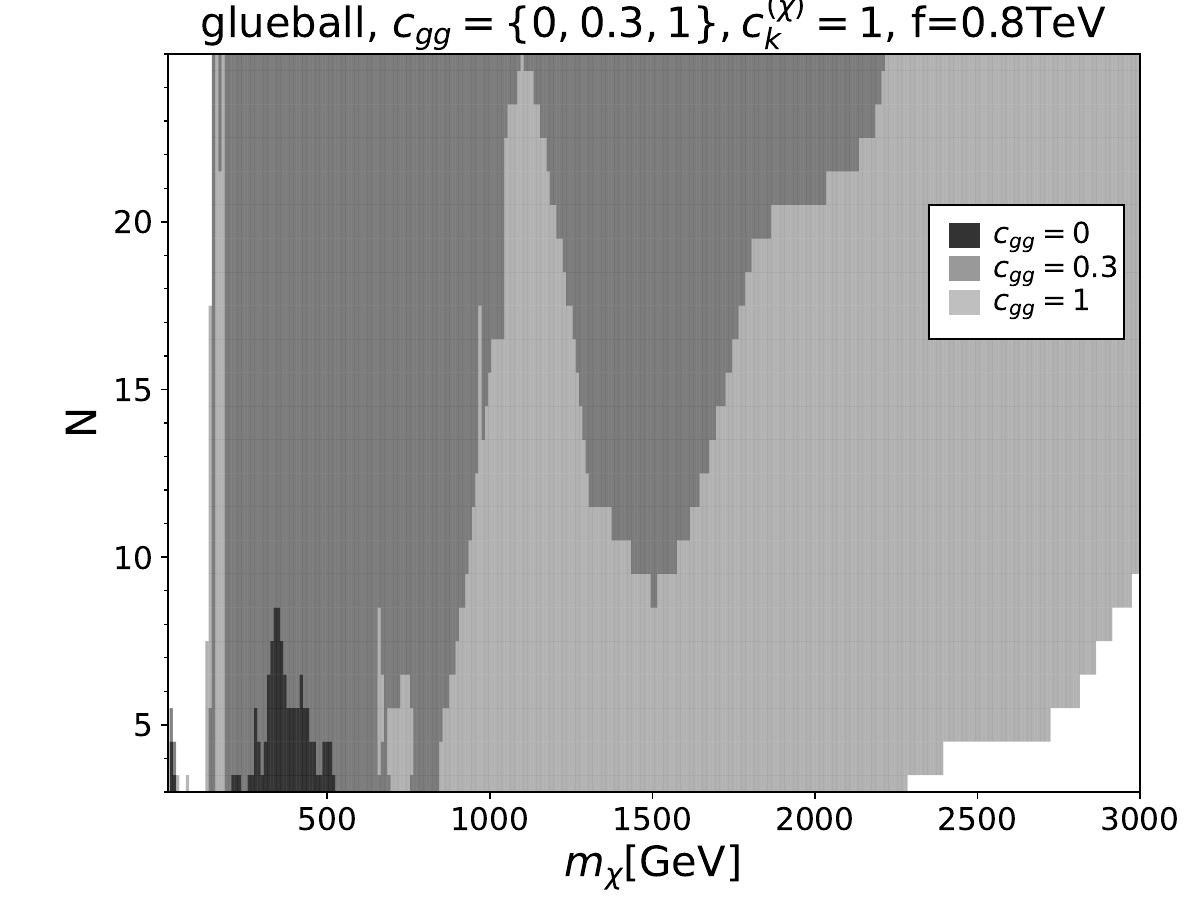}
\hspace{-0.5cm}
\includegraphics[width=0.33\textwidth]{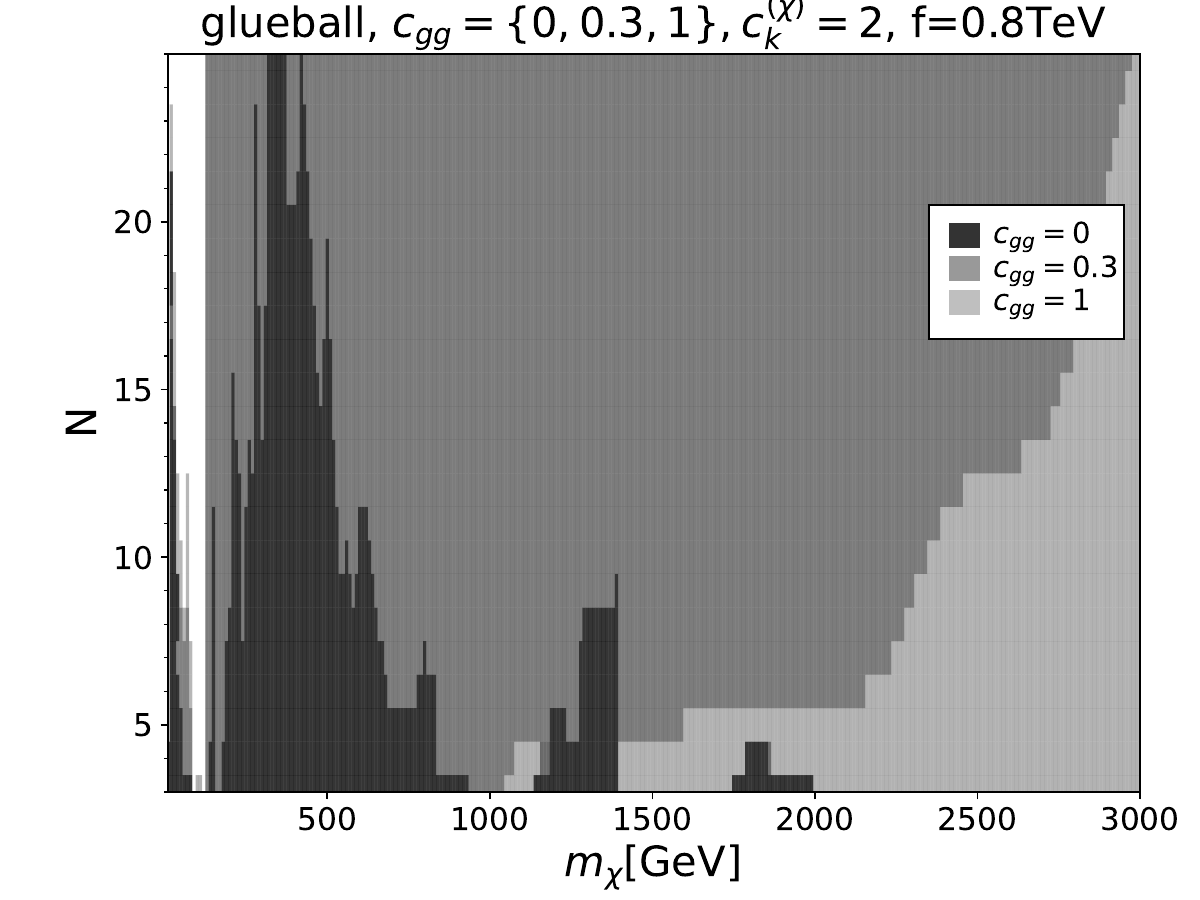}  \\
\rotatebox{90}{\parbox[c]{3.5cm}{\centering \small meson}}
\hspace{-0.25cm}
\includegraphics[width=0.33\textwidth]{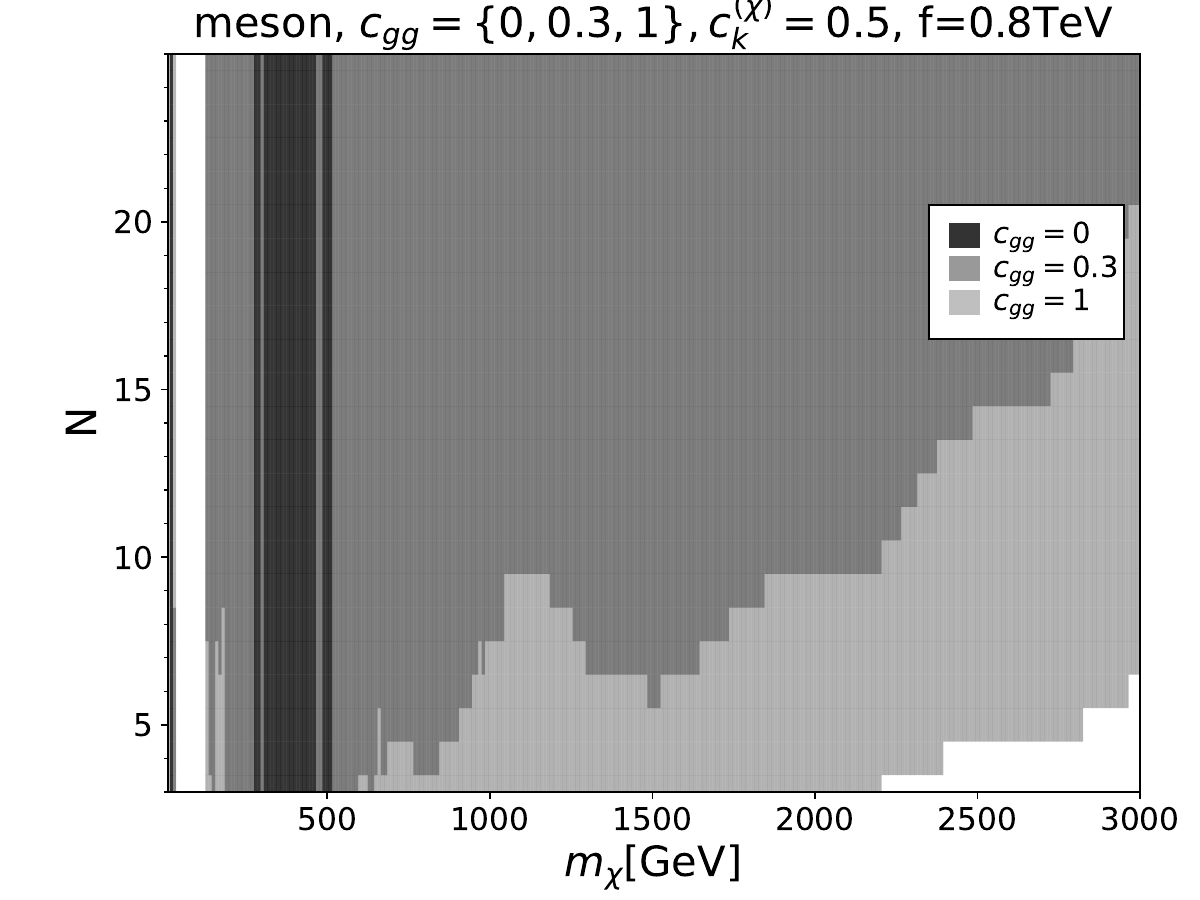}
\hspace{-0.5cm}
\includegraphics[width=0.33\textwidth]{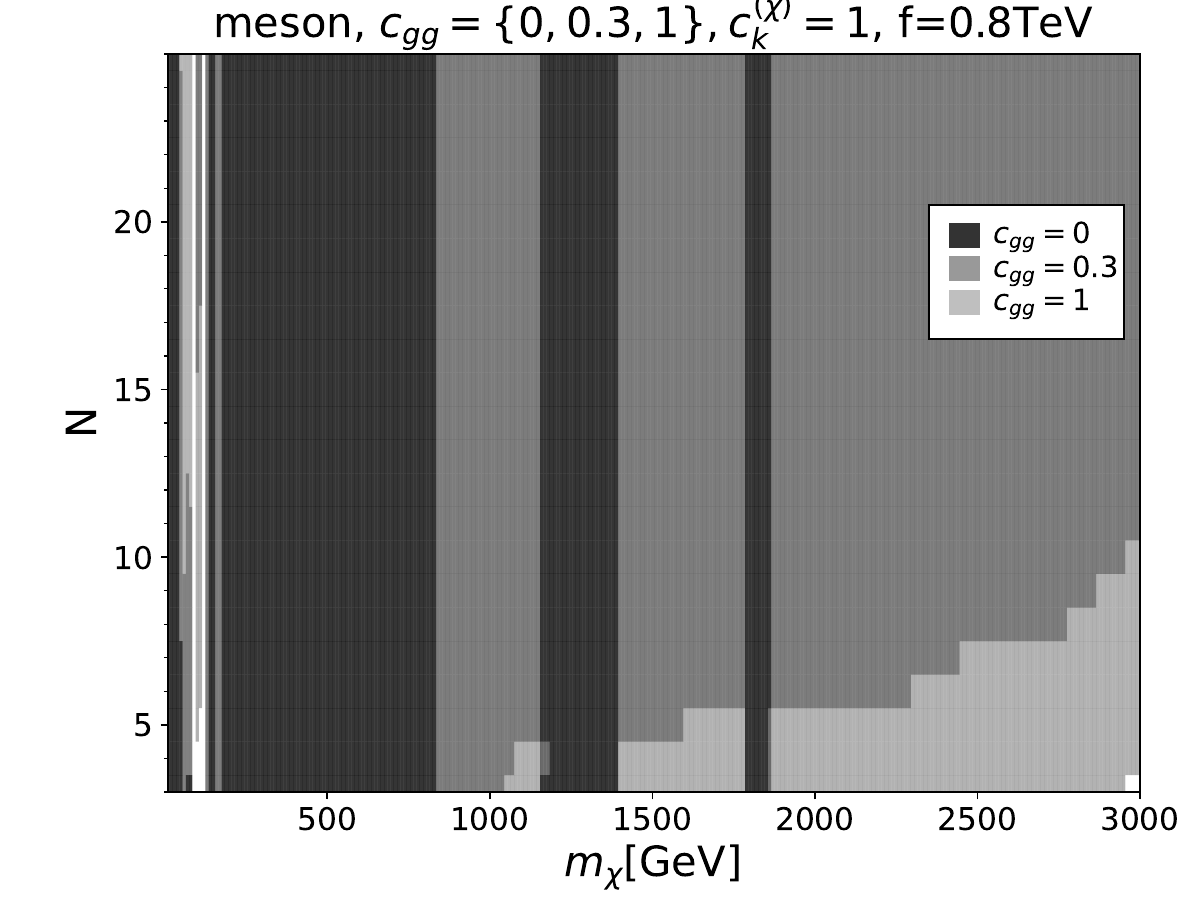}
\hspace{-0.5cm}
\includegraphics[width=0.33\textwidth]{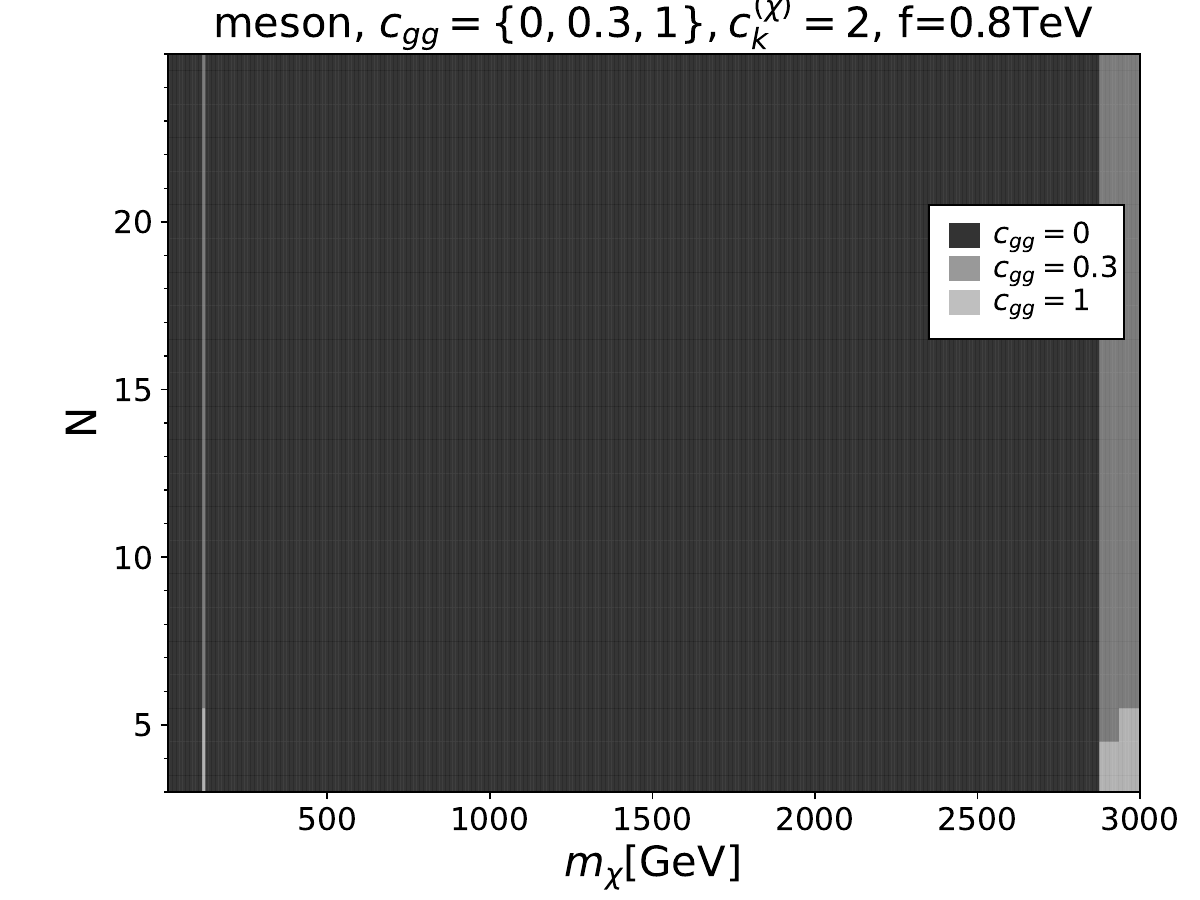} 
\caption{{\it Collider bounds for various choices of $c_k^{(\chi)}$ and $c_{gg}$. Shaded regions are excluded, with darker shades corresponding to lower $c_{gg}$ (defined in Eq.~(\ref{eq:chiGG})). Note that the mixing between the Higgs and the dilaton is very small for large dilaton masses and does not play any role.}}
\label{fig:collider}
\end{figure}

The purple regions in Fig.~\ref{fig:scans_1} are excluded by LHC searches for new scalars~\cite{Bruggisser:2022ofg}, derived using the HiggsTools software~\cite{Djouadi:2018xqq,Bechtle:2020pkv,Bechtle:2020uwn,Bahl:2022igd}. The main coupling controlling dilaton production at the LHC is the contact interaction with gluons generated by the new strong dynamics, 
\be\label{eq:chiGG}
c_{gg} \frac{g_s^2}{3g_*^2} \frac{\chi}{\chi_0} G_{\mu \nu}G^{\mu \nu}, 
\ee
where $c_{gg}$ is an order-one parameter whose exact size depends on the specific UV completion. The darker (lighter) shade of purple for the LHC-excluded regions in Fig.~\ref{fig:scans_1} corresponds to $c_{gg}=0.3$ ($0.5$). Furthermore, in Fig.~\ref{fig:collider} we show the LHC bounds for various choices of the parameters $c_k^{(\chi)}$ and $c_{gg}$, both for the glueball-like and the meson-like $\chi$. The  parameter $c_k^{(\chi)}$ which controls the size of $\chi_0$ suppresses the coupling to gluons, hence it weakens the collider bounds.  As is clear from these plots, the meson case is much more constrained than the glueball one. The reason is that the scale $\chi_0$ suppressing the coupling  of a meson $\chi$ to gluons (\ref{eq:chiGG}) 
is not enhanced by $\sqrt N$ as happens for the glueball $\chi$ according to the relation \eqref{eq:chioff}.
As a result, the combination of the collider bounds with the wrong minimum constraint leaves no viable parameter space for the meson $\chi$, hence we do not discuss this case any further.

Finally, let us comment on the tunnelling angle $h/f$ during the phase transition. In the following section we show that the amount of the CP asymmetry produced during the phase transition can be sensitive to the top quark mass, which would vanish if $h/f = 0$. However, due to SNR effects and the detuning of the Higgs potential at $\chi<\chi_0$ (see the discussion at the end of Appendix~\ref{sec:V1L}) the value of $h/f$ stays of order one during the phase transition.

\section{Electron Electric Dipole Moment}\label{sec:cpv}

We have not yet discussed CP violation in our model. We will briefly present one possible way to introduce it (see also~\cite{Bruggisser:2022rdm}), and demonstrate that the resulting corrections to the electron EDM are substantially suppressed compared to the model without SNR due to the increased dilaton mass.

\begin{figure}
\center
\includegraphics[scale=0.54]{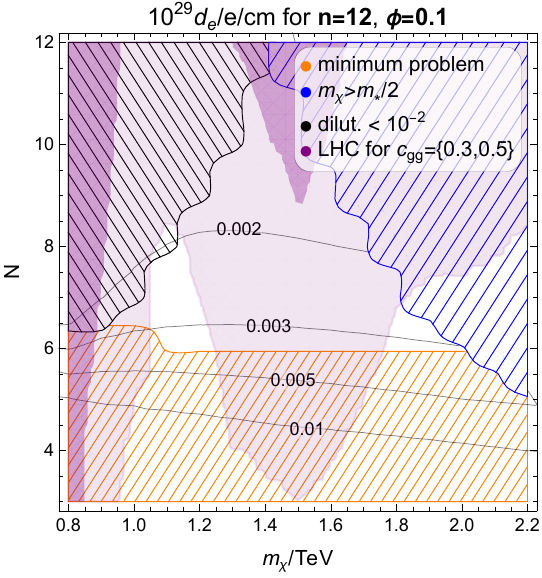}
\hspace{-0.2cm}
\includegraphics[scale=0.54]{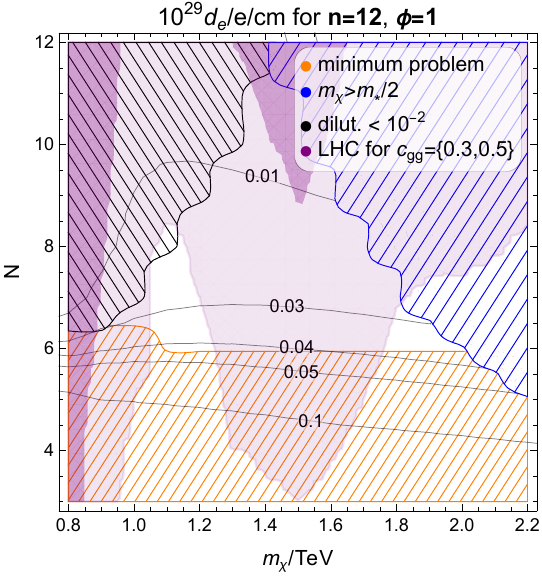}
\hspace{-0.2cm}
\includegraphics[scale=0.54]{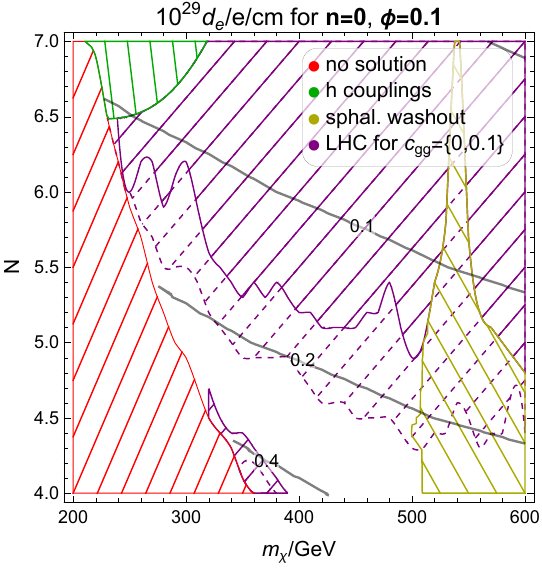}
\caption{{\it Electron EDM (black contours) for a glueball dilaton with parameters as chosen for Fig.~\ref{fig:scans_1}, and $n=12$ (first two plots, for $\phi = 0.1, 1$) and $n=0$ (third plot, $\phi = 0.1$). The color code for the SNR plots is the same as in Fig.~\ref{fig:scans_1}. For the $n=0$ plot  the red hashed region has no viable solutions for the Higgs-dilaton potential, in the yellow-hashed region the baryon asymmetry is washed out by the EW sphalerons after reheating, the purple-hashed (dashed) region is excluded by the LHC dilaton bounds for $c_{gg} = 0$ ($c_{gg} = 0.1$), and the green-hashed region is excluded by constraints on the Higgs couplings.}}
\label{fig:edm}
\end{figure}

We assume that the top quark Yukawa originates from a slightly modified partial compositeness mechanism, where the elementary right-handed top quark couples to two different composite-sector operators ${\cal O}_{L}^{(1)}, {\cal O}_{L}^{(2)}$:
\be
y_{tR}^{(1)} \, \bar t_R {\cal O}_{L}^{(1)}, \quad \; \; y_{tR}^{(2)} \, \bar t_R {\cal O}_{L}^{(2)} \,.
\ee
Below the condensation scale, this results in the top quark Yukawa operator having two contributions:
\be
{\cal L}_{\text {Yuk}}= -\frac{\lambda_t}{\sqrt 2} (g_\chi\chi /g_*) \sin ({h}/{f}) \, \bar t_L t_R
\quad \;\;\text{with}\;\;
\lambda_t = y_{tL} (e^{i \phi} |y_{tR}^{(1)}| + |y_{tR}^{(2)}|) / g_* \,.
\ee
Given that the RG scale which the mixings $y_{tL}$, $y_{tR}^{(1,2)}$ depend on is set by the confinement scale of the new strong sector, the mixings will vary as the dilaton VEV changes during the phase transition. Assuming that a relative complex phase $\phi$ between $y_{tR}^{(1)}$ and $y_{tR}^{(2)}$ exists, the overall phase of the top quark Yukawa will change during the phase transition as long as $y_{tR}^{(1)}$ and $y_{tR}^{(2)}$ scale differently with $\chi$, sourcing the baryon asymmetry~\cite{Bruggisser:2017lhc}. This same relative complex phase also produces CP-violating dilaton and (dilaton-Higgs mixing induced) Higgs interactions with the top quark~\cite{Bruggisser:2018mus, Bruggisser:2018mrt}:
\be
{\cal L}_{\text{Yuk}}  
\, \supset \, - \frac{1}{\sqrt{2}} \left\{\lambda_t+\frac{\partial \lambda_t}{\partial \log \chi} \frac {\chi-\chi_0}{\chi_0} \right\} v_{\rm SM}  \bar t_L t_R \, + \, \text{h.c.}\, ,
\ee
where 
\be
 \text{Im}\left[\frac{1}{\lambda_t}\frac{\partial \lambda_t}{\partial \log \chi}\right] \equiv \text{Im}[\gamma_t]  \propto \phi.
\ee
These CP-violating $h$ and $\chi$ interactions contribute, via two-loop Zee-Barr type diagrams, to the electron EDM. The derivation of the corresponding couplings and all the needed expressions for the computations, utilizing notation identical to the one used in this paper, is given in Section 6 of Ref.~\cite{Bruggisser:2022rdm}. A qualitative understanding of the magnitude of the EDM can be gained from the approximate expression
\be
d_e/e \, \simeq \, 16 \frac{\alpha_{\text{EM}}}{(4 \pi)^3} \sqrt 2 \, G_{\text{F}} \, m_e \,\frac{v_{\text{SM}}}{\chi_0} \,\text{Im}[\gamma_t] 
\left(
 - s_\omega + \frac{m_t^2}{m_\chi^2} \left(\frac{v_{\text{SM}}}{\chi_0} + s_\omega \right)
 \left(1+ \frac 1 3 \log^2 \frac{m_t^2}{m_\chi^2} \right) 
\right) .
\ee
Note that the Higgs-dilaton mixing angle $\omega$ decreases with the mass of the heavier eigenstate, $s_\omega \propto 1/ m_\chi^2$. Hence the overall correction to $d_e/e$ becomes smaller for larger dilaton mass.

In Fig.~\ref{fig:edm}, we show contours of the contributions to the electron EDM, with SNR (first two plots), and without SNR (third plot). In the presence of SNR, for the regions satisfying the collider bounds and constraints on the phase transition temperature, the resulting EDM is two to three orders of magnitude below the current strongest bound~\cite{ACME:2018yjb} which at 90\% CL is given by
\be
|d_e/e| \, < \, 1.1 \cdot 10^{-29} \, \text{cm} \,.
\ee
In the absence of SNR instead, the magnitude of the electron EDM is only a factor of a few lower than the current bound, for the given parameter choice.

\section{Gravitational Waves}\label{sec:grav}

\begin{table}[t]
\centering
\begin{tabular}{|c|c|c|c|c|c| c| }
\hline
& $m_\chi$ & $N$ & $y_{SL}$  & $T_r$& $\alpha$ & $\beta/H[T_r]$ \\
\hline
\text{orange} & $1.3 \,$TeV & $6$ & $1.17$ & $315.2\,$GeV & $6.2$ & $193.1$ \\
\text{cyan} & $1.7 \,$TeV & $7$ & $1.13$ & $390.7\,$GeV & $2.9$ & $276.4$ \\
\text{magenta} & $2 \,$TeV & $6$ & $1.17$ & $439.7\,$GeV & $1.1$ & $459.9$ \\
\hline 
\end{tabular} 
\caption{\it \small Benchmark points for the granvitational-wave spectra, for a glueball dilaton with parameters as chosen for Fig.~\ref{fig:scans_1} and $n=12$. The benchmark points are denoted by dots in Fig.~\ref{fig:scans_1}.}
\label{tab:bmp}
\end{table}

The joined confinement and electroweak phase transition is strongly first-order and is expected to produce a stochastic background of gravitational waves. These can be sourced by sound waves resulting from the expanding bubble walls during the phase transition, by the collision of the bubble walls themselves and by magnetohydrodynamic turbulence which can develop in the plasma. The last contribution is the least well understood but is expected to be subdominant and we will therefore neglect it in the following.

\begin{figure}
\center
\includegraphics[scale=0.56]{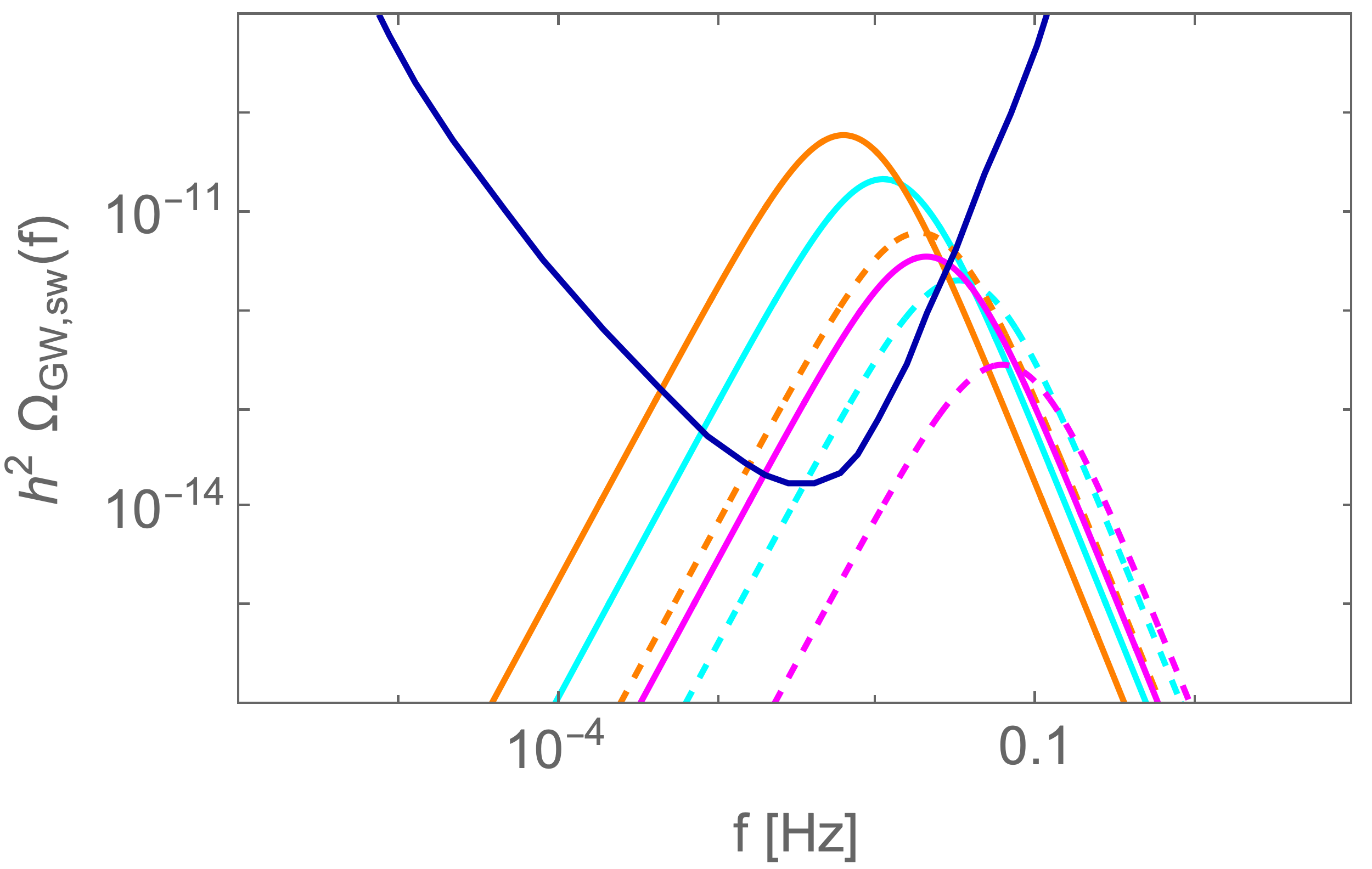}
\hspace{0.3cm}
\includegraphics[scale=0.56]{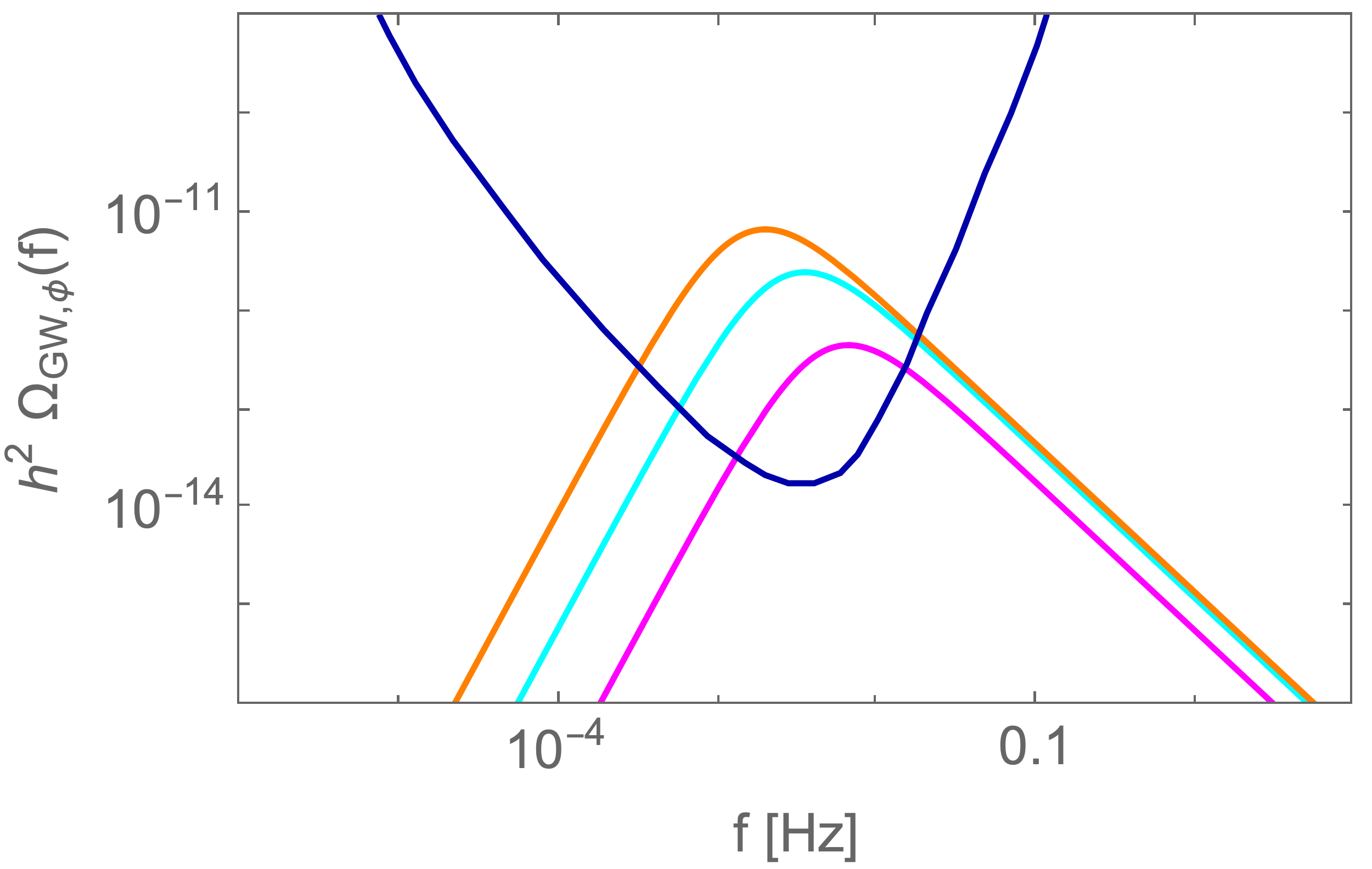}
\caption{{\it Gravitational-wave spectra for the benchmark points given in Table~\ref{tab:bmp}  and denoted by dots in Fig.~\ref{fig:scans_1}, for a glueball dilaton and $n=12$. The other parameters and the color code are as in Fig.~\ref{fig:scans_1}. The considered sources are sound waves (left panel) and bubble wall collisions (right panel) and the assumed bubble wall velocities $v_w=0.9$ (straight lines) and $v_w=0.3$ (dashed lines). We also show the power-law integrated sensitivity curve of LISA as expected for a 3-year mission (blue line).}}
\label{fig:gwspectra}
\end{figure}

The spectra of the produced gravitational waves are mainly controlled by four parameters. The temperature $T_r$ after the phase transition has completed, given in Eq.~\eqref{eq:econs} (see also~Fig.~\ref{fig:scans_2}), determines the amount of redshifting of the spectra from the time of production until today. 
 Furthermore, the parameter
\be
\alpha \, \equiv \, \left(\frac{\Delta V}{\rho_{\rm rad}}\right)_{T_n} 
\ee
measures the strength of the phase transition. Here $\Delta V$ is the free energy difference between the false and the true vacuum and $\rho_{\rm rad}\simeq 3 \pi^2 N^2 T_n^4/8$ is the energy density of the thermal plasma in the false vacuum, with both quantities evaluated at the nucleation temperature $T_n$. Another important quantity is $\beta \equiv [ (d \Gamma / dt) / \Gamma ]_{T_n}$, where $\Gamma$ is the bubble nucleation rate, which measures the inverse duration of the phase transition. Assuming that reheating is fast so that the Hubble rate stays approximately constant, $H[T_n]=H[T_r]$, this gives
\be
\frac{\beta}{H[T_r]}\, \simeq \, \left( T \, \frac{d S_{\rm bnc}}{dT}\right)_{T_n} ,
\ee
where $S_{\rm bnc}$ is the bounce action.
Finally, the spectra also depend on the velocity of the bubble walls $v_w$ during the phase transition. While $T_r, \alpha$ and $\beta$ can be determined straightforwardly, the calculation of the bubble wall velocity is very involved and beyond the scope of this work. In particular, the behaviour of the bubble wall velocity determines the dominant production mechanism of gravitational waves during the phase transition. If the bubble walls enter a runaway regime and the velocity keeps increasing, the dominant source of gravitational waves is bubble wall collisions. In the opposite case, when the bubble walls reach a constant velocity before the phase transition completes, the dominant source is instead sound waves. 
Using the results provided in Ref.~\cite{Baldes:2020kam} we find that the bubble walls do not enter the runaway regime and have moderate velocities $v_w \lesssim 1$. However, a more careful study would be needed to verify the applicability of these results and the approximations made in our specific case. The main reason allowing for slow wall velocities even in our rather supercooled phase transition would be the large number of CFT degrees of freedom acquiring masses upon crossing the bubble wall. We should also stress that moderate values of the wall velocity are crucial for efficient EWBG.
In the following we will simply choose two representative values for the wall velocity, $v_w=0.3$ and $v_w=0.9$.

If the bubble walls do not enter the runaway regime, the dominant source of gravitational waves is thus sound waves.
We use the web-based tool \texttt{PTPlot} \cite{ptplot} to generate the corresponding gravitational-wave spectra which are based on the results from \cite{Hindmarsh:2017gnf,Caprini:2019egz}. In the left panel of Fig.~\ref{fig:gwspectra}, we show the spectra for the three benchmark points given in Table~\ref{tab:bmp} and denoted by dots in Fig.~\ref{fig:scans_1} for the case of a glueball dilaton and $n=12$.  One caveat, though, is that the spectra from sounds waves are only well understood for $\gamma$-factors of the bubble wall $\gamma_w \lesssim 1$, while we cannot exclude that the bubble walls accelerate to ultrarelativisitic velocities given that in our case typically $\alpha > 1$.
The results therefore have to be taken with a grain of salt. In order to get an idea of the range of possible spectra, we also consider the case of gravitational waves produced from bubble wall collisions. In the right panel of Fig.~\ref{fig:gwspectra}, we show the corresponding spectra using the results from \cite{Cutting:2018tjt}, assuming an efficiency factor $\kappa_\phi=1$ and the bubble-wall velocity $v_w=0.9$.
We also plot the power-law integrated sensitivity curve of LISA as expected for a 3-year mission. As one can see, except for the magenta benchmark point with bubble wall velocity $v_w=0.3$, the gravitational waves which are produced for our benchmark points are within reach of LISA.

\begin{figure}
\center
\includegraphics[scale=0.45]{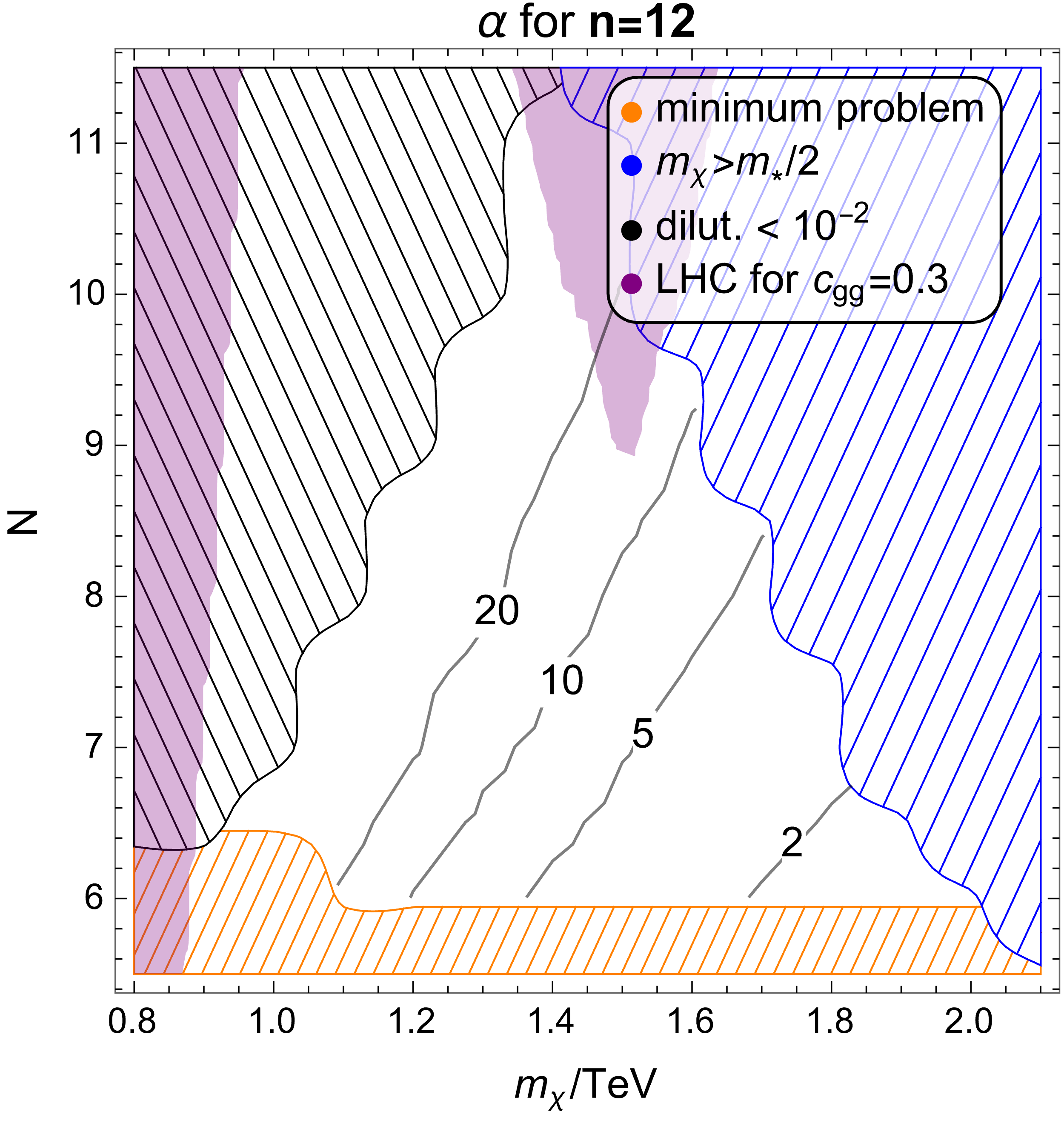}
\hspace{1cm}
\includegraphics[scale=0.45]{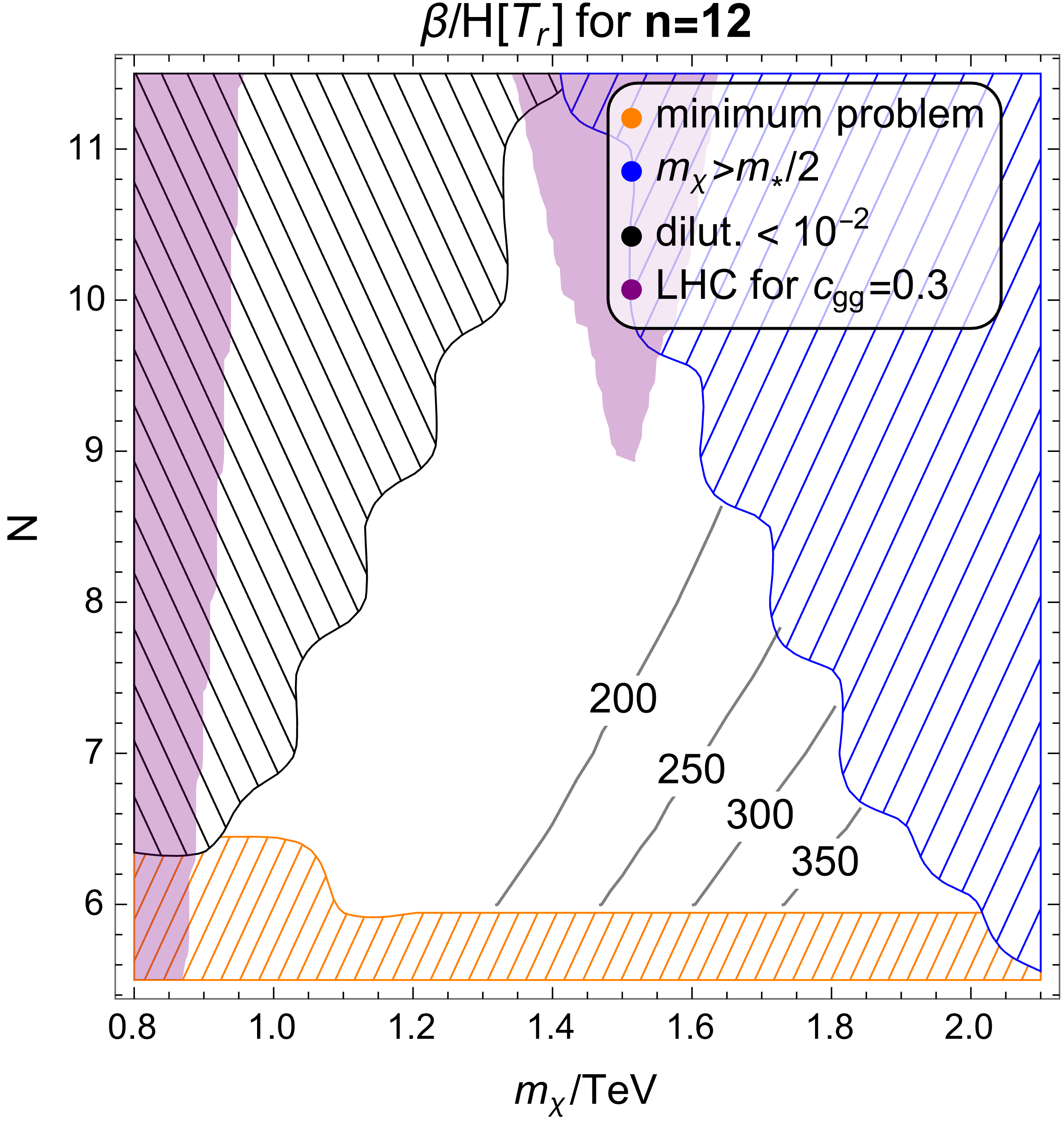}
\caption{{\it Contours (in black) of $\alpha$ and $\beta/H[T_r]$ for a glueball dilaton with parameters as chosen for Fig.~\ref{fig:scans_1} and $n=12$. The color code is the same as in Fig.~\ref{fig:scans_1}. Note that we only show contour lines for $\beta/H[T_r]$ in the lower right corner of the allowed region since in the remaining part the calculation is affected by a numerical instability. It is expected though that $\beta/H[T_r]$ continues to decrease monotonously when going to smaller $m_\chi$ or larger $N$.}} 
\label{fig:alphabeta}
\end{figure}

\section{Discussion} \label{sec:conc}

The minimal composite Higgs framework naturally provides the ingredients necessary for EWBG, such as an EWPT which is first-order and new sources of CP violation. In particular, both can be linked to the composite dilaton field. However, the standard cosmological history of EW symmetry breaking implies that the dilaton 
should  be relatively light to avoid baryon asymmetry washout from 
reheating at the end of the phase transition, leading to tensions with various phenomenological constraints. 
We have implemented a modified dependence of the Higgs VEV on the temperature, relying on the presence of additional singlet fermions in the model. This allows the Higgs VEV to stay large even at large reheating temperatures, and hence permits much larger dilaton masses in the
$800-2000$ GeV range, while the critical and reheat temperatures can reach 400 GeV.

As a result, we find the collider bounds on the dilaton to be easily satisfied in large parts of the parameter space, yet searches in the near future for a heavy dilaton may provide the best way to test this scenario. Additionally, an important test of this scenario can be provided by collider searches for the new composite fermions - the partners of the SNR fermions $S$~\cite{Matsedonskyi:2020kuy}.

The contributions to the electron EDM, linked to EWBG, appear to be highly suppressed with respect to the current experimental sensitivity, hence they may remain unconstraining even after the updates provided by the next generation experiments. A more robust conclusion here, however, requires a detailed study of the baryon asymmetry generation, which is beyond the scope of this paper. In particular, the currently unknown bubble wall velocity during the confinement phase transition can have an important impact on the efficiency of the baryon asymmetry generation and hence on the needed amount of CP violation. However, we point out that the heavy dilaton allowed by SNR tends to decrease the amount of supercooling. This is generally expected to imply slower bubble walls which works in favour of EWBG.

Finally, we found that our scenario of high-temperature EWBG leads to 
sizeable stochastic backgrounds of gravitational waves which are detectable by LISA. However, the dilaton masses on the upper side of the allowed range tend to lead to higher peak gravitational-wave frequencies somewhat away from the best sensitivity region.

The general trend illustrated here 
applies to other EWBG scenarios, i.e.~the addition of new SM singlets 
can enable higher EWBG temperatures~\cite{Meade:2018saz,Baldes:2018nel,Glioti:2018roy,Matsedonskyi:2020mlz,Matsedonskyi:2020kuy,Matsedonskyi:2021hti,Matsedonskyi:2022btb,Biekotter:2022kgf,Chang:2022psj,Bai:2021hfb,Carena:2021onl}, therefore opening the possibility to 
push up the mass of the new field(s) responsible for the first-order phase transition and CP violation,
where LHC and EDM bounds are more relaxed.

\section*{Acknowledgments}

We thank Sebastian Bruggisser for collaboration during the early stage of this project. We also thank Filippo Sala for discussions.
This work is supported by the Deutsche Forschungsgemeinschaft under Germany 
Excellence Strategy - EXC 2121 “Quantum Universe” - 390833306."
OM is supported by STFC HEP Theory Consolidated grant ST/T000694/1. OM also thanks the Mainz Institute for Theoretical Physics (MITP) for its hospitality and support during the completion of this work. BvH would like to thank the ICTP - SAIFR for hospitality during February/March 2023, where part of this work was done, and acknowledges support from FAPESP grant  2021/14335-0.

%\newpage
	
\appendix

\section{Temperature Corrections}\label{sec:temp}

In this Appendix we give details of the thermal corrections to the scalar potential used in our study. For a more detailed explanation and justification of the approximations used we refer the reader to the previous studies, such as~\cite{Creminelli:2001th,Randall:2006py,Nardini:2007me,Konstandin:2010cd,vonHarling:2017yew,Dillon:2017ctw}, and~\cite{Bruggisser:2018mrt} for the specific model that we employ. The one-loop thermal corrections are given by the standard functions
\begin{equation}\label{eq:v1Tloop}
	\Delta V^{\text{1-loop}}_T \, = \, \sum_{\text{bosons}}\frac{n_b  T^4}{2\pi^2}J_b\left[\frac{m^2}{T^2}\right]-\sum_{\text{fermions}}\frac{n_f T^4}{2\pi^2}J_f\left[\frac{m^2}{T^2}\right] ,
\end{equation}
where $n_b$ and $n_f$ are the numbers of degrees of freedom for each species and the functions $J_{b,f}$  are given by
\begin{equation}
J_b[x] \, = \int_0^\infty dk~k^2 \log\left[1-e^{-\sqrt{k^2+x}}\right] \quad \text{and} \quad J_f[x]\, =\int_0^\infty dk~k^2 \log\left[1+e^{-\sqrt{k^2+x}}\right] .
\end{equation}
The sum in Eq.~(\ref{eq:v1Tloop}) includes the SM degrees of freedom, the dilaton and the new SNR fermions. The gauge boson masses are analogous to those in the SM, up to the replacement $h\to (\chi/\chi_0)f \sin h/f$. The only relevant contribution of the SM fermions is that of the top quark with a mass $m_t = (y_t/\sqrt 2) (\chi/\chi_0) f \sin h/f$.

In addition, to model the effect of the CFT degrees of freedom around the origin we add a bosonic contribution with mass equal to $g_\chi \chi$ and multiplicity
\be\label{eq:fixf}
\sum_{\text{CFT bosons}} \hspace{-.2cm} n_{\text{CFT}} \, = \, \frac{45 N^2}{4}.
\ee

\section{One-Loop Effective Potential}\label{sec:V1L}

The presence of a large number of new fermionic states in our model leads to sizeable one-loop corrections to the Higgs and dilaton potential. However, the use of the standard Coleman-Weinberg (CW) expression for the one-loop effective potential in this case turns out to be inadequate for our purposes, as we shall explain below. 

As we have mentioned earlier, the mass of the elementary fermions $m_S^{0}$ is independent of the new strong dynamics scale $\chi$.  During the electroweak phase transition the $\chi$ value changes from zero to $\chi_0$, and therefore there is a region of $\chi$ values where the cutoff of our effective description $m_*$ is below $m_S^{0}$. 
In the limit $m_S^{0} \gg m_*$, a straightforward computation of the CW potential using the expressions for the masses of the $S$ fermions looses sense because it does not account for the presence of new physics at the scales between $m_*$ and $m_S^0$ which can substantially alter the result.  Namely, we expect that the physics at $m_*$ cuts off the momentum integrals and weakens the sensitivity to $m_S^{0}$. In the dual RS picture this screening is realized by an infinite tower of KK resonances with masses $\gtrsim m_*$.

In order to analyze the evolution of the dilaton from zero to $\chi_0$ we therefore need to find a way to interpolate the computation of the quantum corrections between the regimes $m_S^{0}\gg m_*$ and $m_S^{0}\ll m_*$. 

We will start with the usual expression for the one-loop potential before regularization, per one degree of freedom with mass $m$,
\be\label{eq:gen1loop}
V_{\text{1L}} =   \frac {(-1)^F} 2 \int \frac {d^4 k_E}{(2\pi)^4} \ln ({k_E^2+m^2})
\ee
with $F=1 (0)$ for fermions (bosons). We will use the following prescriptions to obtain the potential to use in our study. First, we will add a step-function $\theta (m_*-k)$ to cut off the sensitivity to momenta higher than the compositeness scale $m_* = g_\chi \chi$. This results in
\bea
V_{\text{1L}} 
&=& (-1)^F \frac{1}{64 \pi^2} \left[m_*^4 \left(\ln (m_*^2+m^2) - \frac 1  2\right) + m_*^2 m^2+ m^4 \ln \frac{m^2}{m^2+m_*^2}\right].\label{eq:loopcutoff}
\eea
Furthermore, the one-loop correction to both the Higgs and the dilaton potential is expected to vanish in the absence of elementary-composite interactions, linking the composite sector with the SM states other than the Higgs. In the absence of interactions with an external sector the shift symmetry associated with the Higgs remains unbroken, hence the Higgs should receive no scalar potential. At the same time, when the elementary states decouple from the CFT, all the contributions of the CFT degrees of freedom to the dilaton potential are, by assumption, included in the potential~(\ref{eq:vchi}).   
To account for that we will use the subtracted 1-loop potential
\be\label{eq:vcwbar}
\bar V_{\text{1L}}= V_{\text{1L}}(\{g,g',\lambda_t,y_{SL,SR}\}) - V_{\text{1L}}(\{0,0,0,0\}).
\ee

Decomposing the particle mass $m^2$ into a mixing-dependent part $\delta m^2$ and a mixing-independent one $m_0^2$, we can rewrite the above potential as
\be
\begin{aligned}
\bar V_{\text{1L}} & \, = \, 
\frac {(-1)^F} 2 \int_0^{m_*} \frac {d^4 k_E}{(2\pi)^4} \ln \left[1+\frac{\delta m^2}{k_E^2+m_0^2}\right]\\
&\, = \,
\frac {(-1)^F} 2 \int_0^{m_*} \frac {d^4 k_E}{(2\pi)^4} \frac{\delta m^2}{k_E^2+m_0^2} + \dots,\label{eq:gen1loopbar}
\end{aligned}
\ee
which corresponds to a resummed series of one-loop diagrams with a growing number of mixing insertions. The behaviour of the potential~(\ref{eq:gen1loopbar}) thus closely resembles that obtained for the radion in the 5D dual picture~\cite{Garriga:2002vf}.

It is useful to write down the expressions for the loop corrections in the limits of large or small mixing-independent mass of the loop particle:
\bea
\bar V_{\text{1L}}|_{m_*\gg m} & \simeq & (-1)^F \frac{1}{64 \pi^2} \left(2 \, \delta m^2[\chi]\, m_*^2 +    \dots \right),\label{eq:vbarlim1}\\
\bar V_{\text{1L}}|_{m_*\ll m} &\simeq& (-1)^F \frac{1}{64 \pi^2} \left(  \delta m^2[\chi] \,  \frac {m_*^4 }{m_0^2} + \dots \right). \label{eq:vbarlim2}
\eea

\begin{figure}
\center
\includegraphics[scale=0.8]{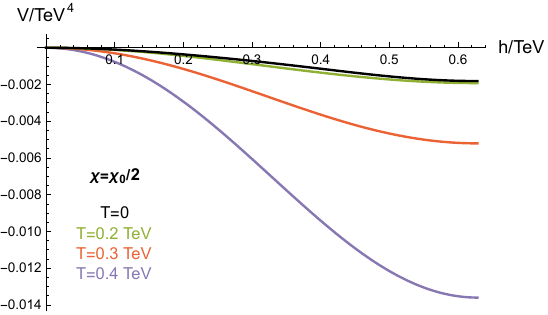}
\caption{{\it Higgs potential at various temperatures, for the glueball dilaton and $n=10$, $y_{SL}=1$, $m_{\chi}=1.5$~TeV, $N=15$, $c^{(\chi)}_k=1$, and the other parameters set as in Table~\ref{tab:bench}. The dilaton VEV equals $\chi_0/2$, where the zero-temperature Higgs potential becomes completely detuned due to the suppression of the loop effects of the new fermions. This results in the minimum of $h$ being displaced to its natural value $h= \pi f/2 (\chi/\chi_0)$.}}
\label{fig:Vdetun}
\end{figure}

We can now discuss one important effect for the phase transition which comes from the one-loop potential. In the minimum today the mass of the new fermions $m_S$ is significantly lower than the cutoff scale $m_*$ and therefore they contribute sizeably to the Higgs potential. As follows from Eq.~(\ref{eq:vbarlim1}), this contribution is $\propto - m_*^2  (m_S^2[h] + m_{\psi}^2[h])$, where we substituted $\delta m_i^2$ with $m_i^2$ which have the same dependence on the Higgs field. 
It is then straightforward to verify that $(m_S^2[h] + m_{\psi}^2[h]) = -(y_{SL}^2+y_{SR}^2)f^2\sin^2 h/f$ up to Higgs-independent terms, and therefore the one-loop correction has a minimum at $h=0$. This contribution has to be finely balanced by the contribution of $\alpha$ in Eq.~(\ref{eq:vCHtree}) in order to push the Higgs VEV slightly away from zero.  
During the phase transition, for $\chi < \chi_0$, the cutoff drops below $m_S$, which results in a relative $(m_*/m_S^0)^2$ suppression of the corresponding loop correction to the Higgs potential (see Eq.~(\ref{eq:vbarlim2})). This in turn destroys the fine-tuning between the different contributions to the potential. The contribution of $\alpha$  becomes dominant and pushes the Higgs VEV all the way to $h=\pi f/2 (\chi/\chi_0)$. 
This is reflected in Fig.~\ref{fig:Vdetun} where we show the zero-temperature Higgs potential for an intermediate value of $\chi$. As a result, the phase transition trajectory becomes shifted to the ``diagonal'' direction $h=\pi f/2 (\chi/\chi_0)$ as shown in Fig.~\ref{fig:sketch}.

\bibliographystyle{JHEP} 
\bibliography{biblio}

\end{document}